\documentclass[11pt,a4paper]{article}
\usepackage{times,helvet}
\usepackage{wasysym}
\usepackage{amsfonts}
\usepackage{accents}
\usepackage{tipa}
\usepackage{changepage}
\usepackage{color}
\usepackage{graphicx}
\usepackage[T1]{fontenc}

\renewcommand{\deg}{^\circ}

\newcommand{\subsubsubsection}[1]{

\vspace{1mm}

\noindent
- \textit{#1}: }

\author{François Vernotte\thanks{F. Vernotte is with UTINAM,
  Observatory THETA of Franche Comt\'{e}-Bourgogne,
  University of Franche Comt\'{e}/UBFC/CNRS,
41 bis avenue de l'observatoire - B.P. 1615,
25010 Besan\c{c}on Cedex - France. Email: \texttt{francois.vernotte@obs-besancon.fr}.}
\hspace{0.5mm} and Satyanad Kichenassamy\thanks{S. Kichenassamy is with the Laboratoire de Math\'{e}matiques de Reims, Moulin de la Housse, B.P.\ 1039, Universit\'{e} de Reims Champagne-Ardenne,
51687 Reims Cedex 2 - France. Email: \texttt{satyanad.kichenassamy@univ-reims.fr}.}%
}

\title{A study of ancient Khmer ephemerides}

\date{\today}

\begin{document}
\maketitle

{% Abstract
\noindent
\it \textbf{Abstract -- }
We study ancient Khmer ephemerides described in 1910 by the French engineer Faraut, in order to determine whether they rely on observations carried out in Cambodia. These ephemerides were found to be of Indian origin and have been adapted for another longitude, most likely in Burma. A method for estimating the date and place where the ephemerides were developed or adapted is described and applied.
}

\section{Introduction\label{sec:intro}}
Our colleague Prof.~Olivier de Bernon, from the \'Ecole Fran\c caise d'Extrême Orient in Paris, pointed out to us the need to understand astronomical systems in Cambodia, as he surmised that astronomical and mathematical ideas from India may have developed there in unexpected ways.\footnote{Prof.~de Bernon made this work possible in many other ways---by bringing about the meeting of its authors in the first place. The remarks on the history of Cambodia and on the Khmer language in this paper are due to him.} A proper discussion of this problem requires an interdisciplinary approach where history, philology and archeology must be supplemented, as we shall see, by an understanding of the evolution of Astronomy and Mathematics up to modern times. This line of thought meets other recent lines of research, on the conceptual evolution of Mathematics, and on the definition and measurement of time, the latter being the main motivation of Indian Astronomy.

In 1910 \cite{faraut1910}, the French engineer Félix Gaspard Faraut (1846--1911) described with great care the method of computing ephemerides in Cambodia used by the \textit{horas}, i.e., the Khmer astronomers/astrologers.\footnote{We are grateful to His Excellency, Mr. Im Borin, Hora of the Royal Palace of Cambodia, for kindly explaining the Khmer system to one of us (F.V.).} The names for the astronomical luminaries as well as the astronomical quantities \cite{faraut1910} clearly show the Indian origin of these methods of computation.\footnote{A system that appears to use the same technical terms is found in Siam, and was discussed by Cassini and de la Loub\`ere, see \cite{la_loubere1691}, \cite[p.~488 sqq.]{jacq-hergoualc'h}.}

Following the works of Billard \cite{billard1971} and Mercier \cite{mercier1985}, we tried to identify the origin of the observations which were necessary to determine the constants of the Khmer ephemerides, i.e. their canon. Thanks to \cite{faraut1910}, we have been able to compute the positions of the astronomical luminaries (Sun, Moon, ascending node of the Moon, Mercury, Venus, Mars, Jupiter and Saturn) and to compare them to the positions given by modern ephemerides \cite{meeus1991, bretagnon_francou1988}.

As shown by Billard with Indian canons \cite{billard1971}, the coincidence of the ephemerides indicates the epoch when the observation of the luminaries were carried out to determine the astronomical constants of the canon. Moreover, Mercier has shown that it is also possible to assess the terrestrial longitude of the location of the place where these observations were conducted \cite{mercier1985}. This yields a method to estimate when and where the Khmer canon constants were determined. The method is described in detail in Section \ref{sec:method}, the results given in Section \ref{sec:results}, and discussed in Section \ref{sec:discussion}. Concluding remarks (Section \ref{sec:conclusion}) close the paper.

Since ancient ephemerides used a sidereal reference whereas modern {ephe\-me\-rides} use a tropical reference, two types of deviations are studied: mean longitude deviations, dominated by the precession of the equinoxes, 
 and synodic deviations, insensitive to them. The fact that both methods indicate the same narrow period provides a validation of both.

\subsection{Relation to earlier work}

Two slightly different scenarios for the constitution of this canon were proposed by Billard, but the papers in which he would have developed his arguments never appeared. In fact, no paper since 1910 appears to even reproduce the data of the system studied by Faraut, let alone analyze them. Billard first suggested \cite[p.~74]{billard1971} that this system was derived from an Indian system with an epoch of 638 AD, that he calls f.638, modified by a longitude correction for a place in Burma, and considered it to be a form of an Indian system, the S\=urya\-siddh\=anta. 
There were several versions of this system, and the latest one seems to have been revised in the light of Brahmagupta's system (seventh century AD);\footnote{According to Prabodhchandra Sengupta's introduction to \cite{burgess1935}, p.~xii: ``Thus from a comparison of astronomical constants [he has] established that there was a book named the S\=urya Siddh\=anta before Var\=aha[mihira]'s time. Var\=aha was one of the first to improve upon it and make it up to date. The present redaction took place decidedly after the time of Brahmagupta.'' On p.~xxiv, we read: ``The \emph{S\=urya Siddh\=anta} has thus undergone progressive changes in its constants and the star table from 400 to 1100 AD. Note also that, according to Alter \cite[p.~281, note 3]{alter}, the translation in \cite{burgess1935} is essentially due to Whitney: ``[a]n initial translation of the \emph{Suryasiddhanta} has been made by Ebenezer Burgess [...]. Yet Whitney, ostensibly serving as Burgess's assistant, was obliged to revise the whole of Burgess's work, which he described as `worthless' " in letters dated March 27, 1858 and October , 1859. It was originally published in \emph{Journal of the American Oriental Society} 6 (1860) 141--498.}
Billard had an earlier one in mind.\footnote{``\textit{Le f.638 qui a fait carrière à travers l'Indochine et dont nous avons, à défaut de l'original sanskrit, les versions birmane, siamoise, laotienne et cambodgienne. Kara\d na du k.(SuryS), il a pour époque dimanche 22 mars 638 AD julien 0h TCUjj ou 1\,365\,702 KYardh . Dans une publication à venir nous pensons être à même de montrer que sitôt son élaboration en Inde, c. 638 AD, ce karana est entré en usage, avec une excellente correction de méridien, en basse Birmanie, d'où, bien plus tard, à partir de la seconde moitié du XIV$^\textrm{\footnotesize e}$ siècle, il a gagné avec le bouddhisme singhalais le Siam, le Laos et le Cambodge. Cette époque de kara\d na explique totalement l'origine de cette ère indochinoise d'époque 638 AD.}''
(``The f.638 which was quite popular throughout %made its way through to
 Indochina and of which we have, in default of the Sanskrit original, Burmese, Siamese, Laotian and Cambodian versions. A Kara\d na of the k.(SuryS) [the canon presented in the \textit{S\={u}rya Siddh\={a}nta}; a \emph{kara\d na} is an astronomical text with a conveniently chosen, contemporaneous epoch], 
it has as epoch Sunday March 22, 638 AD Julian 0h TCUjj [civil time in Ujjain, India] or 1\,365\,702 KYardh [Kaliyuga era starting February 18, 3102 BC at midnight]. In a forthcoming publication we think we are able to show that as soon as it was elaborated in India, i.e. 638 AD, this kara\d na came into use, with an excellent correction of the meridian, in lower Burma, from which, later in the second half of the XIV$^\textrm{\tiny th}$ century, it reached, with Sinhalese Buddhism, Siam, Laos and Cambodia. This epoch of the kara\d na explains entirely the origin of this Indochinese epoch of period 638 AD.''). The four versions mentioned by Billard do not seem to be available.\label{fn:Burma}}
%The f.638 which was quite popular throughout Indochina and of which we have, in lack of the Sanskrit original, the Burmese, Siamese, Laotian and Cambodian versions. Kara\d na of the k.(SuryS) [the canon presented in the \textit{S\={u}rya Siddh\={a}nta}; a \emph{kara\d na} is an astronomical text with a conveniently chosen, contemporaneous epoch], 
%its epoch is Sunday March 22, 638 AD Julian 0h TCUjj [civil time in Ujjain, India] or 1\,365\,702 KYardh [Kaliyuga era starting February 18, 3102 BC at midnight]. In a forthcoming publication we think we are able to show that as soon as it was elaborated in India, ca.~638 AD, this kara\d na came into use, with an excellent correction of the meridian, in lower Burma, from which, much later, from the second half of the XIV$^\textrm{\tiny th}$ century onwards, it gained, with Sinhalese Buddhism, Siam, Laos and Cambodia. This epoch of kara\d na fully explains the origin of this Indochinese era of epoch 638 AD.''). The four versions mentioned by Billard do not seem to be available.\label{fn:Burma}}
(According to the second scenario, suggested in a posthumus paper of his \cite[p.~397]{billard-eade}, the system that found its way into Cambodia was related to an Indian system known as the Parahita system,\footnote{For details on this system, as described by Haridatta, see \cite{parahita}.} probably dated 638 AD.\footnote{``f.638: the \emph{parahita} version of the S\=urya\-siddh\=anta system in sole use in Burma, Thailand and Laos and later used in Cambodia. Outlined in F. G. Faraut, \textit{Astronomie cambodgienne} \dots''}

From another perspective, the Khmer system would appear on the contrary to be quite recent, for the following reasons. First, while the Khmer ephemerides do use an era starting in 638 AD---the \emph{culla-sakar\=aj} or ``small era'' (that Faraut writes ``Chollasakraich")---but this 
era was never used in Cambodia prior to 1848, when it was introduced by King Ang Duong.\footnote{We owe this information to Prof.~Olivier de Bernon, who drew our attention to the problem solved here, and kindly gave us the following details. This ``small era'' appears to have been 
worked out by the Pyu of \'Sr\=\i ksetra. It was adopted by the Burmese when they annexed this kingdom, and then exported to Lanna (Chiang Maï) when it was conquered by the Burmese, then adopted by the Siamese when they took control of Lanna. King Ang Duong had spent thirty years in Bangkok and spoke Siamese.} Also, Faraut's assumptions about the antiquity of the Khmer system were questioned in his obituary by Maître in 1911 \cite{maitre1911}. However, this was 
at a time, prior to C\oe d\`es' work, when the chronology of Cambodia was not firmly 
established, so that his doubts do not necessarily invalidate Faraut's results. We also found that Faraut's data appear to contain inconsistencies in the determination of true longitudes, which is why we deal mostly with mean longitudes.\footnote{One typical inconsistency can be detected in the ``\textit{Grand Chhaya de Mercure}'' \cite[p.~224]{faraut1910} which is a sort of sine table; in column 3, row 2, we find the number $127$ which should be the difference of the numbers in rows 3 and 2, column 2, respectively $474$ and $247$ whose difference is $227$. This mistake is easy to spot but it is much more difficult to see that the number $174$ which is in column 4, row 3 of this table, should be $134$, for trigonometric reasons. %**
 In addition, many calculation errors are widespread throughout this book, such as the one mentioned in footnote \ref{fn:err} p.~\pageref{fn:err}.}

It is therefore necessary to reexamine the evidence. We confirm Billard's suggestions for the most part, but point out small divergences, for some of which we propose an explanation. The Khmer system seems to be much closer to the form of the S\=urya Siddh\=anta that Billard studied, than to the Parahita system as described by Haridatta, that incorporates corrections dated 684 AD.\footnote{Perhaps Billard and Eade had a different version of the Parahita system in mind.} Therefore, despite their late date, documents in this case do contain material that goes back to the seventh century. The question whether this system was imported into Cambodia only in the nineteenth century, or whether it was present much earlier, went into disuse, and was reintroduced in Cambodia in the nineteenth century cannot be answered with the information at our disposal.

To enable comparison with Billard's results, we followed his procedure and notation throughout, taking advantage, however, of Mercier's reinterpretation of the method as a nonlinear least squares fitting method. We modified it by taking into account a more recent model for the modern ephemerides and for the Earth's rotation. We now turn to possible drawbacks of Billard's approach, and how they are avoided here.

\subsection{Billard's approach}

Indian or Khmer texts do not furnish tables of observations, but predictive mathematical models. Major texts present themselves as emendations or restorations of older canons that have become inconsistent with observation. Therefore, a comparison with modern data could narrow down rather precisely the time of composition of the canon. Indeed, it is
 reasonable to expect that any model would be roughly accurate around the time of the composition of the text that introduces it; otherwise, the most casual observation, of eclipses for instance, would prove it false. Billard \cite{billard1971} argued that the time (if unique) where the difference between the mean positions given by an ancient theory and the modern one would be least possible could be an estimate of the time of the observations that supported this theory. He applied this approach systematically to a variety of (mostly Indian) systems. However, his method was not widely adopted for a number of reasons.

First, he interpreted variations from modern positions as errors of observations that would be distributed normally. Since the major treatises of this period describe astronomical instruments \cite{ohashi,sr-sarma}, and claim consistency with observation, it is reasonable to expect that systematic error had been recognized and taken into account. The assumption of normality would then be reasonable if a large number of measurements had been made with the same apparatus. At the same time, Billard postulated \cite[\S 2, 1, 14]{billard1971}, ``a \emph{single} series of astronomical observations, \emph{all of them contemporaneous [with one 
another]}, and very narrowly clustered about a central epoch $T$.''\footnote{``une série 
\textit{unique} d'observations astronomiques  \textit{toutes contemporaines}, et très étroitement groupées autour d'une époque centrale $T$'' (emphasis is Billard's).}

What if only one measurement was made? How do we know that errors on different luminaries may be treated as independent random variables with the same law? Is this law necessarily normal? Mercier \cite{mercier1985} worked around this issue by reinterpreting the method as a least-squares fitting method, rather than an analysis of measurement errors, and showed its usefulness in other contexts \cite{mercier}.

Second, Billard worked on the mean positions of the luminaries, whereas the measured quantities would correspond to the true ones.\footnote{Mercier \cite{mercier2005} shows on significant examples that the analysis does extend to true longitudes, with similar results.} For this 
reason, we che\-cked our results using true positions, finding a result consistent with the one obtained via mean longitudes. Unfortunately, Faraut's data seem to be corrupt, and we had to 
correct what appear to be obvious errors. For this reason only, we have relegated this analysis to an appendix, and worked in most of the paper with mean longitudes.

Third, Billard assumed that angles were the measured quantities, and Pingree pointed out that ``\textit{[t]here were no adequate instruments for measuring celestial angles}'' (see \cite{pingree1976} p. 116).\footnote{Pingree's thesis in this paper seems to have been influenced not only by an inadequate appreciation of the conceptual framework of ancient Indian Mathematics, but also by a faulty reading of primary sources. See \cite{ohashi,mak,filliozat,mercier2005}.} Now, Indian works do not mention measurements of angles because they do not use the very notions of angle or parallel at all; Indian Geometry has developed other mathematical tools that make them unnecessary \cite{sk-2010,sk-2012}. The texts do not describe the evolution of celestial angles, but of arc-lengths on special circles of various radii, either directly or through the product of their sines or cosines by the appropriate radius. Roundoff and conversion from sine to arc must be taken into account in estimating accuracy. The notion of angle, as a magnitude attached to the meeting of two lines is never mentioned, and it is not a ``primary'' notion that would have to be part of any moderately sophisticated mathematical theory. This last point was clarified only in the last two centuries; the modern (``Bourbaki'') point of view in Mathematics is that the measurement of an angle necessarily relies on the rectification of an arc of a circle, and cannot be achieved by ``elementary'' means. Another difference with Hellenistic Mathematics is the admission of a variable unit of length, leading to a \emph{scale-calculus}, that seems to be 
the only known way to account for the earliest Indian rule for the quadrature of the circle extant \cite{sk-2006}. The value chosen for the radius of the ``trigonometric circle'' may influence roundoff procedures. A proper statistical model for errors in observation must be based on the quantities actually measured, and on the conceptual framework that underlies the \emph{modus operandi}. Of course, late works directly influenced by Hellenistic texts could evince knowledge of the notion of  angle; this may even serve as a \emph{shibboleth} of foreign influence.

For these reasons, we have treated, following Mercier, deviations in longitudes as fitting errors between ancient and modern systems rather than measurement errors. Whatever the latter may have been, the existence of a rather narrowly identifiable epoch where the deviations for all luminaries are simultaneously small suggests that actual observations were carried out around this time. %**

\section{Methods\label{sec:method}}
In this work, we used the notation of \cite{billard1971}.

	\subsection{Choice of the astronomical luminaries}
We focused this study on 8 astronomical luminaries: the Sun, the Moon, the lunar ascending node, Mercury, Venus, Mars, Jupiter and Saturn (see Table \ref{tab:luminaries}). The ascending node of the Moon plays an important role since it was considered as a celestial body in its own right, named Rea Hou in Cambodia, derived from Sanskrit \textit{R\={a}hu}, responsible for eclipses.\footnote{ The opposite point is called \emph{Ketu} in Sanskrit. \textit{R\={a}hu} was adopted by Arabic Astronomy and, later, by Western Astronomy as the well-known ``head of the dragon'' (\textit{caput draconis}), \textit{Ketu} being its ``tail'' (\textit{cauda draconis}). 
It was very recently pointed out \cite{bryan} that it was incorporated into the Arthurian legend by Geoffrey of Monmouth in the twelfth century, through a new etymology of the name of Arthur's father Uther Pendragon. Bryan notes that this object was taken to be a comet by Wace; oddly enough, \textit{Ketu} may also refer to a comet in Sanskrit. For the transmission of Indian Astronomy through Arabic authors, see chapter VII of \cite{mercier}.} 
 To these luminaries, we have added the Vernal point. However, the lunar apogee, which was studied in \cite{billard1971}, has not been used in the present work.

\begin{table}
\hspace{-18mm}\begin{tabular}{|l|c|c|c|c|c|c|c|c|c|c|}
%\begin{tabular}{|l|p{1cm}|p{1cm}|p{1cm}|p{1cm}|p{1cm}|p{1cm}|p{1cm}|p{1cm}|p{1cm}|p{1cm}}
\hline
Luminary & Vernal & Sun & Moon & Lunar & Lunar & Mercury & Venus & Mars & Jupiter & Saturn\\
 & point & & & apogee & asc. node & & & & & \\
\hline
Symbol & \vernal & \astrosun & \leftmoon & $\varpi$ & $\theta$ %\ascnode
 & \mercury & \venus & \mars & \jupiter & \saturn \\
\hline
$i$ & 1 & 2 & 3 & 4 & 5 & 6 & 7 & 8 & 9 & 10\\
\hline
\end{tabular}
\caption{Symbols and numbers used to designate luminaries. In this study, the lunar apogee was not used, but we kept it for the sake of compatibility with \cite{billard1971}.\label{tab:luminaries}}
\end{table}

	\subsection{Choice of the astronomical quantities}
		\subsubsection{Mean longitude deviations}
Following the works that Billard \cite{billard1971} and Mercier \cite{mercier1985} performed for Indian ephe\-me\-rides, we have simply calculated the mean longitudes rather than the true longitudes. The use of the true longitudes, that we undertook at the beginning of this study 
(see Appendix), gives results that are in perfect agreement with those obtained with the mean longitudes but more dubious insofar as they are very sensitive to the residuals errors that remain in \cite{faraut1910}. Denoting $\mathcal{L}_i(t)$ the Khmer mean longitude in degrees of 
luminary $i$ (see the value of $i$ in Table \ref{tab:luminaries}) at instant $t$ and $L_i(t)$ the modern mean longitude in degrees of luminary $i$ at instant $t$, we define the mean longitude deviation $X_i(t)$ as their difference:
\begin{equation}
X_i(t)=\mathcal{L}_i(t)-L_i(t).
\end{equation}
The Khmer mean longitude are referred to a fixed sidereal point which is the Vernal point of the epoch when the astronomical constants of the ephemerides were determined\footnote{Around 500 AD, the Vernal point was close to $\zeta$ Piscium, a relatively faint star (magnitude 5.3). As Biot rightly points out (see \cite{biot1862}, p.16), the ancient astronomers could not directly refer to such a star to measure practically the luminary longitudes. A secondary reference star, much brighter, should have been used. But we know no more about this secondary reference star than about the procedure that was followed.}. On the other hand, modern ephemerides are referred to the current Vernal point. Because of the precession of the equinoxes, both mean longitude systems will converge only for the date when the fixed sidereal reference coincided actually with the Vernal point, i.e. the epoch when the astronomical constants of the ephemerides were determined.\footnote{According to Faraut \cite[pages 75 and 79]{faraut1910}, Khmers used to take the precession of the equinoxes into account, and argues that one rule may be accounted in this way. We have insufficient information to determine which of the medieval theories of precession (or libration) of the equinoxes, if any, is involved here. For these theories, see \cite[Ch. II]{mercier}.} Therefore, this is an accurate method to estimate this very date.

		\subsubsection{Synodic deviations\label{sec:syno_dev}}
On the other hand, we define the synodic longitudes $\mathcal{L}_{\astrosun i}(t)$ and $L_{\astrosun i}(t)$ as the mean longitudes referenced to the Sun:
\begin{equation}
\left\{\begin{array}{l}
\mathcal{L}_{\astrosun i}(t)=\mathcal{L}_i(t)-\mathcal{L}_2(t)\\
L_{\astrosun i}(t)=L_i(t)-L_2(t)
\end{array}\right.
\end{equation}
where $\mathcal{L}_2(t)$ and $L_2(t)$ are the mean longitude of the Sun respectively in the Khmer and the modern ephemerides. The synodic deviation $X_{\astrosun i}(t)$ is then defined as:
\begin{equation}
X_{\astrosun i}(t)=\mathcal{L}_{\astrosun i}(t)-L_{\astrosun i}(t).
\end{equation}
Since the reference is now the position of the Sun in both systems, the synodic deviation will no longer be affected by the precession of the equinoxes. 
 However, if the astronomical constants determined by the ancient Khmer astronomers are slightly erroneous, we should observe a very slow drift between both synodic longitudes. 
 This yields another method to estimate the epoch when the astronomical constants of the ephemerides were determined. 
 This second method is entirely independent of the previous one.

It may be noticed that the position of the Vernal point being 0 in each system, we can compute the difference of the references of the Khmer and modern systems, i.e. the effect of the precession of the equinoxes on the Sun, by:
\begin{equation}
X_{\astrosun 1}=\left[0-\mathcal{L}_2(t)\right]-\left[0-L_2(t)\right]=L_2(t)-\mathcal{L}_2(t)=-X_2(t)\label{eq:gamma}.
\end{equation}
%(**c'est bien $X_{\astrosun 2}$ ?) Non, vous avez raison c'est $X_{\astrosun 1}$ !

	\subsection{Khmer ephemerides}
		\subsubsection{Khmer reference time\label{sec:Khmer_reference_time}}
For Khmer horas, the time origin (day 0 of year 0) was March 21, 638 AD at midnight, beginning of the \textit{Cullasakar\=aj} 
(era)\footnote{Actually, only the first day of year 1 \textit{Cullasakar\=aj}, which is March 22, 639 AD, is defined in the ancient texts. 
 Note that this epoch differs from that mentioned by Billard by precisely one year but the concept of day 0 of year 0 is nothing but a modern fancy to facilitate the conversion into Julian days! %***
The two would be compatible if one referred to the elapsed year as opposed to the current year. If not, we must conclude that Billard did not make use of Faraut's data.} (Julian Day 1\,954\,167.5 plus the time lag between Greenwich and Cambodia, i.e $\sim$7 h).

As mentioned by Faraut \cite[p.~13]{faraut1910}, the astronomical day begins at midnight. However, he claims later on that the longitudes of all luminaries are determined %for
 at sunrise \cite[p.~132]{faraut1910}, i.e. approximately at 6 AM since
Cambodia is not very far from the equator ($\sim 11\deg$ North). Our results are quite incompatible with this later time convention (unless considering a meridian 90$\deg$ West of Cambodia!) and show clearly that the ephemerides are always given for 0 AM. We will adopt this convention in this study.

		\subsubsection{Khmer mean longitude model}
For the Khmer ephemerides, the mean longitude $\mathcal{L}_i$ expressed in degrees of planet $i$ at instant $t$ is given by a linear relationship:
\begin{equation}
\mathcal{L}_i(t)=\alpha_i t + \beta_i\label{eq:Khmer_mean_longitude}
\end{equation}
where $t$ is the time, i.e. the number of days elapsed since the time origin, $\alpha_i$ is the mean motion of planet $i$ in $\deg/$day and $\beta_i$ is the mean longitude in degrees of planet $i$ at the time origin of Khmer ephemerides. 
The constants $\{\alpha_i\}$ and $\{\beta_i\}$ are given for each luminary in Table \ref{tab:alpha_beta}. It may be noticed that these constants are always given as ratios of integer numbers since all computations were carried out with integers.
 These ratios were obtained from the ``recipes'' given by Faraut in \cite{faraut1910}, which are a list of additions, subtractions, multiplications, divisions with remainder and quotient involving only integer numbers. Each of these procedures were replaced by a single floating-point operation.

For information, we give below the example of Mars. Page 214 of \cite{faraut1910}, Faraut explains:

\begin{adjustwidth}{20pt}{20pt}
``{\it Le temps que cet Astre met à parcourir les douze Réasseys\footnote{\textbf{Reasseys:} (Sanskrit: \textit{r\=a\'si}) Arc of 30$\deg$ corresponding to a constellation of the zodiac. While we give the standard Sanskrit equivalents for the main technical terms, it will be seen in the comments on these terms that transmission from Sanskrit to Khmer is not the only possibility. For this reason, some terms from other languages that are known to have had influence in Cambodia are indicated, without aiming at completeness. Further terms, some of which have obvious Indian counterparts, may be found in Faraut's glossary \cite[pp.1--4]{faraut1910}.}, mois solaires, appelé son Ch\oe ung Ha, son diviseur, est de $687$ jours. Il représente son année.

On peut remarquer qu'il est le même que celui trouvé par les Astronomes Européens.

Avec ce nombre on détermine tout d'abord le Phol Langsak\footnote{\textbf{Langsak:} Astronomical landmark day at the beginning of a year. One recognizes behind \textit{phol}---that means fruit, result, according to Faraut---the Sanskrit \textit{phala}, with the same meanings (cf.~Tamil \textit{pa\b lam} ``fruit etc.'', \textit{pala\b n} ``proceeds, influence, etc.'' or \textit{paya\b n} ``results etc.''). Similarly, one may surmise that \textit{sak} is related to \textit{sakar\=aj} (era) and hence to Sanskrit \textit{\'saka}, that may mean era (cf. Tamil \textit{cak\=aptam} = \textit{\'saka$+$abda}). As for \textit{lang}, it might be related to Khmer \textit{\d lae\.n} (to rise, to augment) or to Sanskrit \textit{lagna} (ascendant), even though the treatment in Khmer of voiced consonants in foreign words is not in favor of the second possibility. These hypotheses require confirmation, all the more since Faraut's transliteration is not always faithful, as the term ``Chollasakraich'' for \textit{Cullasakar\=aj} illustrates.} [\ldots].

Il s'obtient en ajoutant $633$ au Harakoune\footnote{\textbf{Harakoune:} Number of elapsed days since the beginning of the era. This corresponds to Sanskrit: \textit{aharga\d na}; this word appears to have reached Cambodia through Thailand, because McFarland's \textit{Thai English Dictionary} (p.~916) gives \textit{aragu\d na} with the same meaning.} $454\,018$ dont la somme $454\,651$ est divisée par $687$ et donne $661 + 544$. Le reste $544$ de cette division est le Phol.

Il faut, maintenant, chercher le Mathiouma\footnote{\textbf{Mathiouma:} (Sanskrit: \textit{madhyama}) Mean longitude of a luminary.} du jour Langsak [\ldots].

À cet effet, le Phol $544$ est multiplié par $12 = 6\,528$, ce produit est divisé par le Ch\oe ung Ha $687$ d'où $6528 : 687 = 9 + 335$\footnote{\label{fn:err}The notation $a : b=q+r$ means that $a=bq+r$. Actually, the correct remainder is $345$.}. 
 Le quotient $9$ représente les Réasseys et le reste $335$ est multiplié par $30 = 10\,050$. Ce produit est encore divisé par $687$ soit $10\,050 : 687 = 11 + 432$. 
 Le quotient $11$ représente le nombre d'Angsas\footnote{\textbf{Angsas:} (Sanskrit: \textit{a\d m\'sa}) $1/30$ Reassey $=1\deg$.} et le reste $432$ est multiplié par $60$, ce qui donne $25\,020$, produit que l'on divise encore par $687$, soit donc : $25920 : 687 = 37 + 501$.

Le quotient $37$, toujours augmenté de $7$ unités $= 44$, représente les Lipdas\footnote{\textbf{Lipdas:} (Sanskrit: \textit{lipta}) $1/60$ Angsa $=1'$.} et le reste $501$ s'appelle le Pouichalip.

Le Mathiouma de Mars est donc : R.$9$ A.$14$ L.$44$.}''\footnote{``The time this Planet takes to traverse the twelve Reasseys, the solar month, called its Ch\oe ung Ha, its divisor, is $687$ days. It represents its year.

It may be noted that it is the same one as that found by the European Astronomers.

With this number we first determine the Phol Langsak [\ldots].

It is obtained by adding $633$ to the Harakoune $454\,018$ whose sum $454\,651$ is divided by $687$ and gives $661 + 544$. The remainder $544$ of this division is the Phol.

We must now look for the Mathiouma of the day Langsak [\ldots].

For this purpose, the Phol $544$ is multiplied by $12 = 6\,528$, this product is divided by the Ch\oe ung Ha $687$, whence $6\,528 : 687 = 9 + 335$. The quotient $9$ represents the Reasseys and the rest $335$ is multiplied by $30 = 10\,050$. This product is further divided by $687$, i.e. $10\,050 : 687 = 11 + 432$. The quotient $11$ represents the number of Angsas and the remainder $432$ is multiplied by $60$, giving $25\,020$, which is further divided by $687$, that is to say: $25920 : 687 = 37 + 501$.

The quotient $37$, always increased by $7$ units $= 44$, represents the Lipdas and the remainder $501$ is called the Pouichalip.

The Mathiouma of Mars is thus: R.$9$ A.$14$ L.$44$.''}.
\end{adjustwidth}

This text means that for $t=454\,018$ days since the beginning of the \textit{Cullasakar\=aj} era, i.e. Langsak of 1243 \textit{Cullasakar\=aj} $=$ April 14$^\textrm{\footnotesize th}$, 1881, the mean longitude (Mathiouma) of Mars is $\mathcal{L}(t)=12\times (t+633)/687$ in Reasseys. 
 Since $1$ Reassey $= 30\deg$, this yields the following constants: $\mathcal{L}(t)=\alpha_8\times t + \beta_8$ with $\alpha_8=30\times 12/687=360/687=120/229\deg$/d and $\beta_8=30\times 12\times 633/687=75960/229\deg$. 
 The other operations are intended to give a result between 0 and 12 Reasseys and to convert the decimal part into Angsas ($\deg$) and Lipdas (').

However, a small correction is introduced at the end of the process: ``\textit{Le quotient [\ldots], \textbf{toujours augmenté de $7$ unités} [\ldots], représente les Lipdas}''. Thus, a $7'$ angle is systematically added to the computed mean longitude of Mars. This correction appears separately from the $\beta_8$ coefficient in Table \ref{tab:alpha_beta}. Its possible role will be discussed in \S \ref{sec:dif_beta}.

The same evaluation of the $\{\alpha_i\}$ and $\{\beta_i\}$ coefficients, including an eventual correction, were carried out for all luminaries and are reported in Table \ref{tab:alpha_beta}.

It may be noticed that La Loubère and Cassini \cite{la_loubere1691} gives exactly the same ``recipe'' for computing the mean longitude of the Sun and the Moon with the same corrective terms than \cite{faraut1910}, i.e. 3' for the Sun and 40' for the Moon (see \cite{la_loubere1691}, Tome second, Règles de l'astronomie, \S IV--VII for the Sun and \S X for the Moon).

\begin{table}
\hspace{-15mm}\begin{tabular}{|l||c|r|r||c|r|}
\hline
 & $\alpha_i$ & \multicolumn{1}{c|}{$360/\alpha_i$} & \multicolumn{1}{c||}{Modern} & $\beta_i$ & \multicolumn{1}{c|}{$\beta_i$} \\
Luminary & ratio & \multicolumn{1}{c|}{decimal} & \multicolumn{1}{c||}{period} & ratio & \multicolumn{1}{c|}{decimal} \\
 & ($\deg/$day) & \multicolumn{1}{c|}{(days)} & \multicolumn{1}{c||}{(days)} & ($\deg$) & \multicolumn{1}{c|}{($\deg$)}\\
\hline
\hline
&&&&&\\
Sun & $\displaystyle\frac{288\,000}{292\,207}$ & $365.2588$ & $365.2422$ & $\displaystyle-\frac{1\,119}{2\,435}-\frac{3}{60}$ & $-0.51$ \\
&&&&&\\
\hline
&&&&&\\
Moon & $\displaystyle\frac{21\,090}{1\,730}+\frac{288\,000}{292\,207}$ & $27.32167$ & $27.32158$ & $\displaystyle\frac{4\,554\,663}{421\,255}-\frac{40}{60}$ & $10.15$ \\
&&&&&\\
\hline
&&&&&\\
Lunar asc. node & $\displaystyle-\frac{8}{151}$ & $-6\,795.000$ & $-6\,798.384$ & $\displaystyle\frac{27\,520}{151}$ & $-177.75$ \\
&&&&&\\
\hline
&&&&&\\
Mercury & $\displaystyle\frac{36\,000}{8797}$ & $87.97000$ & $87.96843$ & $\displaystyle\frac{2\,007\,720}{8\,797} - \frac{1}{60}$ & $-131.79$ \\
&&&&&\\
\hline
&&&&&\\
Venus & $\displaystyle\frac{1\,200}{749}$ & $224.7000$ & $224.6954$ & $\displaystyle\frac{253\,080}{749} - \frac{2}{60}$ & $-22.14$ \\
&&&&&\\
\hline
&&&&&\\
Mars & $\displaystyle\frac{120}{229}$ & $687.0000$ & $686.9297$ & $\displaystyle\frac{75\,960}{229} + \frac{7}{60}$ & $-28.18$ \\
&&&&&\\
\hline
&&&&&\\
Jupiter & $\displaystyle\frac{1\,080}{12\,997}$ & $4\,332.333$ & $4\,330.596$ & $\displaystyle\frac{1\,100\,160}{12\,997} - \frac{1}{60}$ & $84.63$ \\
&&&&&\\
\hline
&&&&&\\
Saturn & $\displaystyle\frac{180}{5\,383}$ & $10\,766.00$ & $10\,746.94$ & $\displaystyle\frac{235\,980}{769}$ & $-53.13$ \\
&&&&&\\
\hline
\end{tabular}
\caption{Constants $\{\alpha_i\}$ and $\{\beta_i\}$ of equation (\ref{eq:Khmer_mean_longitude}) given by \cite{faraut1910}. Rather to give the decimal values of the $\{\alpha_i\}$, we give the corresponding revolution periods, i.e. $\{360/\alpha_i\}$ (column 3) and we compare them to the modern values of the revolution periods (column 4). It may be noticed that mean motion of the Moon is given relatively to the Sun and the one of the Sun must be added. In column 6, the decimal values of $\{\beta_i\}$ are given within $\left]-180\deg,+180\deg \right]$. In this column, we add another term introduced without explanation by Faraut at the $\beta_i$ of the Sun, the Moon, Mercury, Venus, Mars and Jupiter. %** (référence exacte ?) %respectively $-3'$ and $-40'$ 
(see \cite{faraut1910}, pp. 30, 35--36, 222, 234, 214 and 229 respectively for the Sun, the Moon, Mercury, Venus, Mars and Jupiter).
 A possible interpretation will be given in \S \ref{sec:dif_beta}.\label{tab:alpha_beta}}
\end{table}

	\subsection{Modern ephemerides}
		\subsubsection{Modern mean longitude model}
For the Sun, the Moon, the lunar ascending node, Mercury, Venus and Mars, we used a simple model which takes into account secular terms as far as $t^4$ \cite{meeus1991}. Since Jupiter and Saturn are affected by periodic mutual resonances, we used a much more sophisticated model, VSOP87 \cite{bretagnon_francou1988}, including a large number of trigonometric terms besides a development up to $t^5$.

		\subsubsection{Modern reference time}
			\subsubsubsection{Slowing down of the Earth rotation}
For accurate ephemerides over a long period (here 2500 years), the slowing down of the Earth rotation must be taken into account. Whereas Mercier used the Spencer Jones formula \cite{spencer_jones1939}, we preferred a more recent model given by Morrison and Stephenson \cite{morrison_stephenson1982}:
\begin{equation}
\Delta t = \textrm{TT} - \textrm{UT} =-15 + \frac{(\textrm{JD} - 2\,382\,148)^2}{41\,048\,480}\label{eq:slowing_down_Earth}
\end{equation}
where $\Delta t$, TT and UT are expressed in seconds, TT is the Terrestrial Dynamical Time, UT the Universal Time and JD the Julian Day. This model ensures an error of less than 20 minutes over the last 2500 years.

			\subsubsubsection{Terrestrial longitude of the place of observation}
The time argument used in modern ephemerides is the Terrestrial Time, TT, which is the Universal Time UT corrected from the slowing down of the Earth thanks to equation (\ref{eq:slowing_down_Earth}). UT is the time of the meridian origin, Greenwich. Therefore, the longitude of the astronomical observatory must be taken into account since it yields a time lag.

For the sake of simplicity as well as neutrality with our main hypothesis, we assume a location at 90$\deg$ East, i.e. between the prime meridian of ancient Indian ephemerides, Ujjain (76$\deg$ East), and Phnom Penh (105$\deg$ East). This corresponds
 to a time advance of
 $90/360=0.25$ days or 6 h
 relative to the Greenwich meridian time. Therefore, at 0 AM on day $j$ at
 $90\deg$ East, it is 6 PM on day $j-1$ at Greenwich. The argument $t$ used in Khmer Ephemerides should be corrected in $t_G$ at Greenwhich as:
\begin{equation}
t_G=t-0.25\label{eq:timelag}
\end{equation}
where $t$ and $t_G$ are expressed in days.

However, the exact terrestrial longitude of the astronomical observatory will be a parameter of this study and a corrective term $\Delta \phi_0$ will be estimated
 (see \S \ref{sec:observatory_longitude_estimation})
 in such a way that the exact location where observations were performed will be:
\begin{equation}
\phi_0=90\deg+\Delta \phi_0.
\end{equation}

	\subsection{Time scales}
As stated above, TT is the time scale for modern ephemerides. It is generally expressed in Julian Days, JD, which are decimal numbers whose decimal part gives the time of day. The origin of JD, i.e. JD$=0$, is assigned to the day starting at noon on January 1$^\textrm{\footnotesize st}$, 4713 BC. The change of day occurs at noon at the Greenwich meridian. %(TT).
However, the time reference $t_K$ of Khmer ephemerides are days of \textit{Cullasakar\=aj} Era and the change of day occurs at 0 AM. We must also take into account the time lag due to our assumed location at 90$\deg$ East longitude.

 The difference between JD and \textit{Cullasakar\=aj} origins is then 1\,954\,167 days (see \S \ref{sec:Khmer_reference_time})
 plus the time between the beginning of this Julian Day (noon) and the beginning of the \textit{Cullasakar\=aj} day (midnight) $+0.5$ day,
 plus the $-0.25$ day longitude time lag of equation (\ref{eq:timelag}), %difference between Greenwich and 90$\deg$ East (negatively counted to the East) $-90/360=-0.25$ day,
 so in total $+0.25$ days. The effect of slowing down of the rotation of the Earth modeled by (\ref{eq:slowing_down_Earth}) must also be taken into account.

Thus, the conversion relationship between these two time scales is:
\begin{equation}
% In the code:
%	merid_0=-90; 						// Central terrestrial longitude of the study
%	h_0=0; 							// Time reference for the Khmer ephemerides
%	dif_jj_hrkn=1954167; 					// Origin difference between Julian Days and Chollasakraich era
%	cor_ral=db(2715.6)+db(573.36)*TC+db(46.5)*TC*TC; 	// Slowing down of the Earth
%	cor_ral/=db(86400); 					// Conversion from seconds to days
% 	TT=db(jj)+db(0.5)+h_0/db(24)+merid_0/db(360)+cor_ral;	// Time for modern ephemerides (in days)
% 	TK=db(jj-dif_jj_hrkn);					// Time for Khmer ephemerides (in days)
%\left\{
%\begin{array}{lcr}
\textrm{TT}  =  t_K + 1\,954\,167 + 0.25 + \displaystyle\frac{\Delta T}{86\,400}%\\
%&&\\
%t_K & = & \textrm{TT} - 1\,954\,167 - 0.25 - \displaystyle\frac{\Delta T}{86\,400}\\
%\end{array}
%\right.
\label{eq:convert_TT_tK}
\end{equation}
where $\Delta T$ is expressed in seconds and 86\,400 is the number of seconds in a day.

We will introduce below an additional term, $\Delta\phi$, which will represent the terrestrial longitude error.

	\subsection{Assessment of the observation epoch and longitude\label{sec:observatory_longitude_estimation}}
In the following, the choice of the luminaries used to assess the observation epoch and the observatory longitude will be of importance. According to \cite{billard1971} and \cite{mercier1985}, the choice of a luminary set $I$ will be indicated by a sequence of 10 1's or 0's, indicating whether a particular value of $i$ defined in Table \ref{tab:luminaries} is included or not. For instance, (10101\,01101)\label{sec:notations} indicates that the Vernal point, the Moon, the lunar ascending node, Venus, Mars and Saturn are selected but neither the Sun, the lunar apogee, Mercury nor Jupiter.
		\subsubsection{Direct method on the mean longitude deviations\label{sec:direct_method}}
Since the Khmer mean longitudes of luminaries follow a simple linear model, it's quite simple to introduce the new parameter $\Delta \phi$ of which the optimal value $\Delta \hat\phi_0$ will be our estimate of $\Delta \phi_0$:
\begin{equation}
\mathcal{L}_i(t,\Delta\phi)=\alpha_i t+\frac{\alpha_i}{360}\Delta \phi + \beta_i=\mathcal{L}_i(t)+\frac{\alpha_i}{360}\Delta \phi.
\end{equation}
Thus, the mean longitude deviation is given by:
\begin{equation}
X_i(t,\Delta\phi)=\alpha_i t+\frac{\alpha_i}{360}\Delta \phi + \beta_i - L_i(t)=X_i(t)+\frac{\alpha_i}{360}\Delta \phi.\label{eq:Xk_DeltaPhi}
\end{equation}
Let us remind the reader %** 
 that with this sign convention, $\Delta \phi$ is negatively counted towards the East.

Neglecting the higher order terms in the modern mean longitudes, we can assume that they are linear versus time in the neighborhood of $t_0$ for each luminary:
\begin{equation}
L_i(t)=A_i t +B_i.
\end{equation}
Therefore, $X_i(t)$ becomes:
\begin{equation}
X_i(t)=(\alpha_i-A_i) t + \beta_i -B_i=a_i t+ b_i\label{eq:first_step}
\end{equation}
where $a_i=\alpha_i-A_i$ and $b_i=\beta_i -B_i$
and then (\ref{eq:Xk_DeltaPhi}):
\begin{equation}
X_i(t,\Delta\phi)=a_i t+ b_i+\frac{\alpha_i}{360}\Delta \phi.
\end{equation}

The main assumption of this method consists in considering that there exists a value pair $(\hat{t}_0,\Delta\hat{\phi}_0)$ which minimizes the mean longitude deviations of each luminary:
\begin{equation}
X_i(\hat{t}_0,\Delta\hat{\phi}_0)=a_i \hat{t_0}+b_i+\frac{\alpha_i}{360}\Delta \hat{\phi}_0\approx 0.
\end{equation}
Therefore, we can obtain the estimates $(\hat{t}_0,\Delta \hat{\phi}_0)$ of the true values $(t_0,\Delta \phi_0)$ by solving the following system for all the $n$ selected luminaries $i$:
\begin{equation}
\left\{\begin{array}{c}
\vdots \\
b_i=-a_i \hat{t}_0 -\displaystyle\frac{\alpha_i}{360}\Delta \hat{\phi}_0.\\
\vdots
\end{array}
\right.\label{eq:system_t0_DPhi0}
\end{equation}
Denoting $B$ the $n\times 1$ column vector of the $b_i$ and $P$ the $2\times 1$ column vector of the parameter estimates $\hat{t}_0$ and $\Delta\hat{\phi}_0$, it is possible to rewrite (\ref{eq:system_t0_DPhi0}) in matrix form:
\begin{equation}
B=A\cdot P
\end{equation}
where $A$ is the $n \times 2$ following matrix:
\begin{equation}
A=\left(\begin{array}{cc}
\vdots & \vdots\\
-a_i & -\displaystyle\frac{\alpha_i}{360}\\
\vdots & \vdots
\end{array}\right).
\end{equation}
The residuals are then $R=B-A\cdot P$ and their variance is $S=R^T R$. 
 From $S$, the uncertainties over $\hat{t}_0$ and $\Delta \hat{\phi}_0$ may be assessed by using the classical relationships giving the variances $\sigma_t^2$, $\sigma_{\Delta\phi}^2$ 
and their covariance $\textrm{Cov}_{t,\Delta\Phi}$, by assuming that the residuals follow a normal 
 distribution\footnote{This point is arguable since long-term correlations between the residuals may appear, but this assumption gives an order of magnitude.}:
\begin{equation}
Cv=\displaystyle \frac{1}{n-2}S(A^TA)^{-1}=\left(\begin{array}{cc}
\sigma_t^2 & \textrm{Cov}_{t,\Delta\Phi}\\
\textrm{Cov}_{t,\Delta\Phi} & \sigma_{\Delta\phi}^2
\end{array}\right).
\end{equation}
Confidence intervals over $t_0$ and $\Delta \phi_0$ may then be computed by taking into account that $(\hat{t}_0-t_0)/\sigma_t$ and $(\hat{\Delta}\phi_0-\Delta \phi_0)/\sigma_{\Delta\phi}$ follow a Student distribution with $n-2$ degrees of freedom.

We propose then a method relying on four steps.

			\subsubsubsection{First step}
We compute the longitude deviations for all luminaries and for date varying from 500 BC to 2000 AD by step of 8 days.

			\subsubsubsection{Second step}
We perform a first set of linear regressions over (\ref{eq:first_step}) to assess the pair $(a_i, b_i)$ for each luminary.

			\subsubsubsection{Third step}
We solve the system (\ref{eq:system_t0_DPhi0}) by using least squares %**
 to obtain an estimation of $(t_0,\Delta \phi_0)$.

			\subsubsubsection{Fourth step}
We calculate the uncertainties over $(t_0, \Delta\phi_0)$ and express the result as a confidence interval.

Steps 3 and 4 can be repeated for different luminary sets.

		\subsubsection{Variance method on the synodic deviations\label{sec:variance_method}}
The synodic deviations are based upon the positions of the planets and the Moon relative to the Sun. As it has already been stated (see \S \ref{sec:syno_dev}), this provides a way to evaluate the quality of the ephemerides independently of the Vernal point and, correlatively, of the precession of the equinoxes. The synodic longitudes provided by the Khmer ephemerides should be very close to the ones determined by the modern ephemerides for the epoch corresponding to the observations realized for defining the Khmer ephemeride constants $\{\alpha_i,\beta_i\}$. For other epochs, we should observe an increasing deviation due to the errors in the $\alpha_i$ constants.

Following the approach of Billard \cite{billard1971} and Mercier \cite{mercier1985}, we processed the synodic deviations by using a method based upon the minimum variance which provides at once an estimation of both parameters $(t_0,\Delta \phi_0)$.

Using the notation of Mercier (see \cite{mercier1985} p. 99), the empirical variance %(** on peut distinguer explicitement variance empirique et variance de la distribution---ou ne pas insister sur ce point) FV: Pour expliciter sur ce point, il faudrait donner un modèle de distribution ainsi que les raisons qui nous poussent à le choisir, ce qui risque de nous emmener assez loin...
 of the synodic deviations of a set $I$ of luminaries for a given date $t$ is:
\begin{equation}
Q=\frac{1}{n-1} \sum_{i\in I} \left[X_{\astrosun i}(t)-\bar{X}_{\astrosun}(t)\right]^2\label{eq:variance}
\end{equation}
where $n$ is the number of luminaries in the set $I$, i.e. the number of 1's in the chosen $I$, and $\bar{X}_{\astrosun}(t)$ is the mean of the set of $\left\{X_{\astrosun i}(t)\right\}$ at date $t$:
\begin{equation}
\bar{X}_{\astrosun}(t)=\frac{1}{n}\sum_{i\in I}X_{\astrosun i}(t).
\end{equation}

\begin{figure}
\includegraphics[width=\columnwidth]{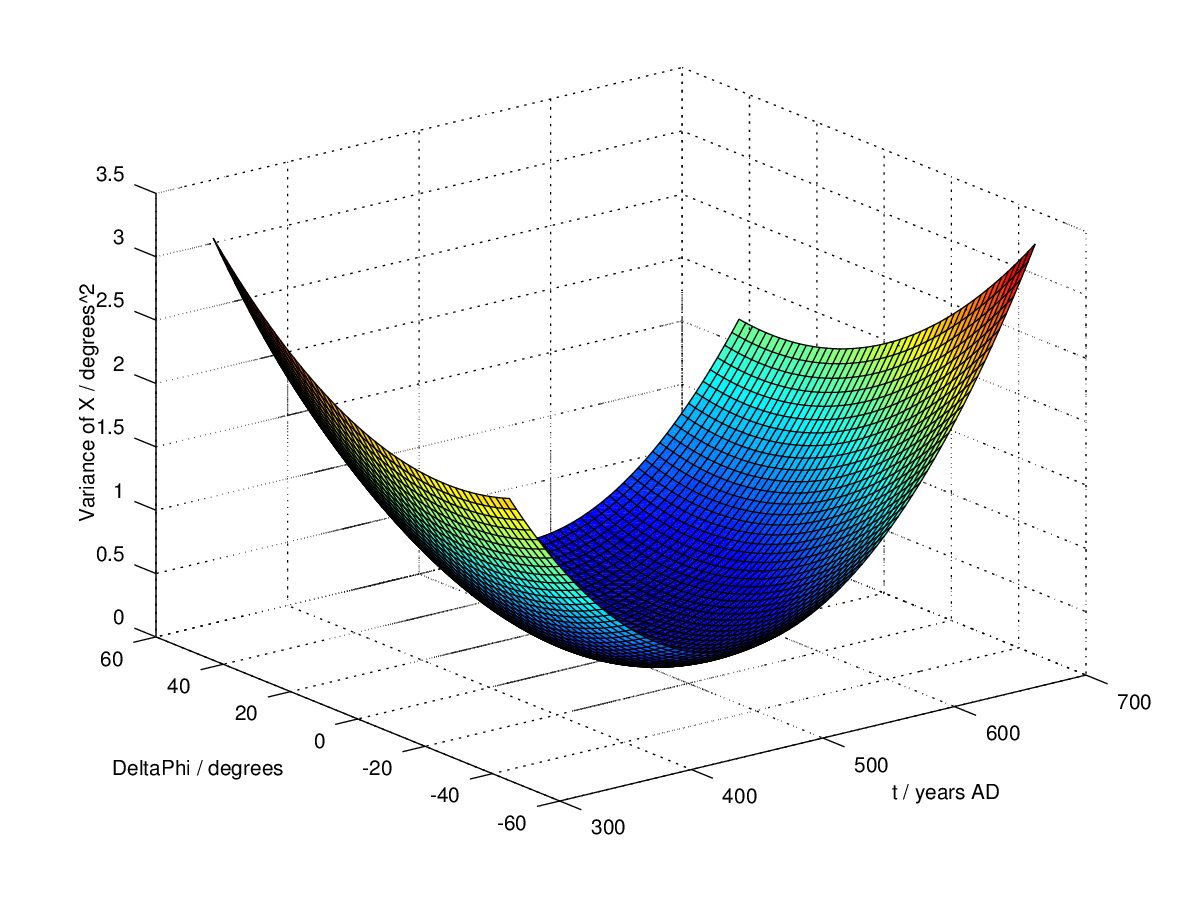}
\caption{Example of 3-D plot of the variance of the synodic deviations versus $t$ and $\Delta\phi$ using the luminaries (10101\,01100) (see \S \ref{sec:notations} page \pageref{sec:notations}). This surface is very close to a paraboloid.\label{fig:quasi_paraboloid}}
\end{figure}

Versus $t$ and $\Delta \phi$, the variance $Q$ may be approximated as a paraboloid (see Figure \ref{fig:quasi_paraboloid}) in the neighborhood of its minimum $(t_0,\Delta \phi_0)$:
\begin{equation}
Q(t,\Delta\phi) \approx h_{11}(t-t_0)^2 + 2 h_{12}(t-t_0)(\Delta\phi-\Delta\phi_0)+ h_{22}(\Delta\phi-\Delta\phi_0)^2+ Q_0\label{eq:nonlinear}
\end{equation}
where $h_{11}$, $h_{12}$, $h_{22}$ and $Q_0$ are the other parameters of the paraboloid to be estimated. Therefore, these are nuisance parameters while $t_0$ and $\Delta\phi_0$ are our parameters of interest. It must be highlighted that (\ref{eq:nonlinear}) is a nonlinear relationship of the parameters and of the quantities %**
 $Q$, $t$ and $\Delta\phi$. Therefore, the estimates $\{\hat{t}_0,\Delta\hat{\phi}_0,\hat{h}_{11}, \hat{h}_{12}, \hat{h}_{22},\hat{Q}_0\}$ of the true parameters $\{t_0,\Delta\phi_0,h_{11}, h_{12}, h_{22},Q_0\}$ can be obtained by using a nonlinear least squares algorithm (we used the Gauss-Newton algorithm, see for instance \cite[pp. 163--191]{hansen_et_al2013}).

According to Mercier (see \cite{mercier1985} p. 99), the variance of the estimate $\hat{t}_0$ and $\Delta\hat{\phi}_0$ are given by\footnote{We replaced $Q_0$ by $|Q_0|$ since the fit of this parameter may provide a negative value.}:
\begin{equation}
\left\{\begin{array}{rcl}
\sigma_t^2&=&\displaystyle \frac{|Q_0|}{n-r} \quad \frac{h_{22}}{h_{11}h_{22}-h_{12}^2}\\
\sigma_{\Delta\phi}^2&=&\displaystyle \frac{|Q_0|}{n-r} \quad \frac{h_{11}}{h_{11}h_{22}-h_{12}^2}
\end{array}\right.
\end{equation}
where $n$ is the number of random variables in each $Q$, i.e. the number of luminaries in the set $I$, %and $r=1$
minus 1
 because we determine the variance estimates $Q$ in (\ref{eq:variance}) by estimating the mathematical expectation $E\left[X_{\astrosun}(t)\right]$ from the arithmetic mean $\bar{X}_{\astrosun}(t)$, and $r$ is the number of parameter of interest, i.e. $r=2$. Similarly, we deduce that the covariance between the estimates $\hat{t}_0$ and $\Delta\hat{\phi}_0$ is given by:
\begin{equation}
\textrm{Cov}_{t,\Delta\phi}=\displaystyle \frac{|Q_0|}{n-r} \quad \frac{h_{12}}{h_{11}h_{22}-h_{12}^2}.
\end{equation}
%Denoting $(\hat{t}_0,\Delta\hat{\phi}_0)$ the estimates of the true values $(t_0,\Delta \phi_0)$, s
Since $(\hat{t}_0-t_0)/\sigma_t$ and $(\hat{\Delta}\phi_0-\Delta \phi_0)/\sigma_{\Delta\phi}$ follow a Student distribution with $n-r$ degrees of freedom, it is easy to define a confidence interval either on $t_0$, $\Delta\phi_0$ or the couple $(t_0, \Delta\phi_0)$.

For this method, we propose then another four step process.
			\subsubsubsection{First step}
We compute the synodic deviations for all luminaries, for date varying from 300 AD to 700 AD by step of 100 days and for $\Delta \phi$ varying from $-45\deg$ to $+45\deg$ by step of $1/4\deg$ around a central longitude $\phi_c=90\deg$ East.
			\subsubsubsection{Second step}
We compute the variances $Q$ according with (\ref{eq:variance}).
			\subsubsubsection{Third step}
We estimate the parameters of interest $(t_0, \Delta\phi_0)$ as well as the nuisance parameters $(h_{11}, h_{12}, h_{22}, Q_0)$ by applying the nonlinear least squares method to (\ref{eq:nonlinear}).
			\subsubsubsection{Fourth step}
We calculate the uncertainties over $(t_0, \Delta\phi_0)$ and express the result as a confidence interval.

Steps 2, 3 and 4 can be repeated for different luminary sets.

\section{Results\label{sec:results}}

	\subsection{Preliminary results}
\begin{figure}
\hspace{-6.5mm} \includegraphics[height=7.6cm]{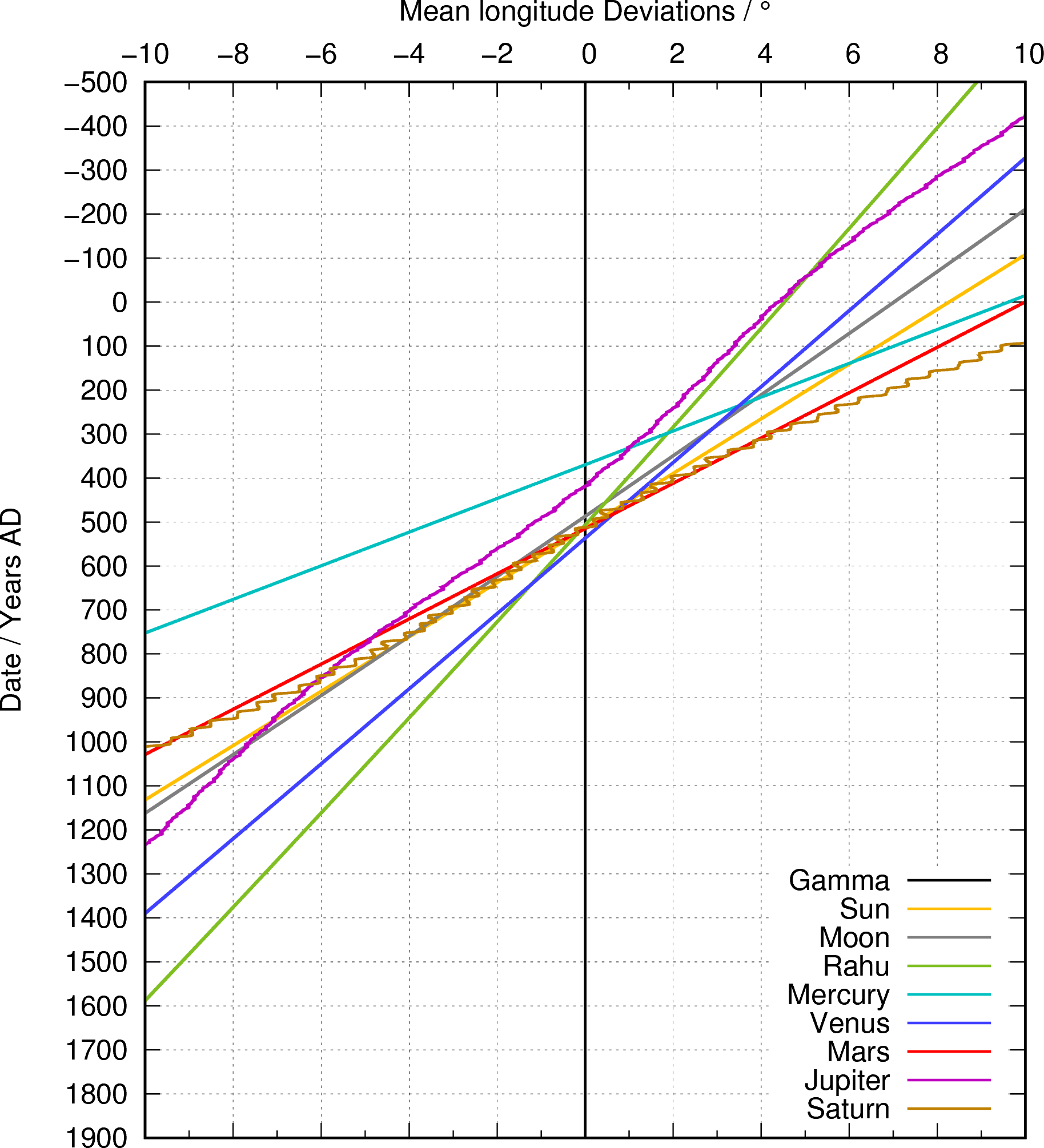} \includegraphics[height=7.6cm]{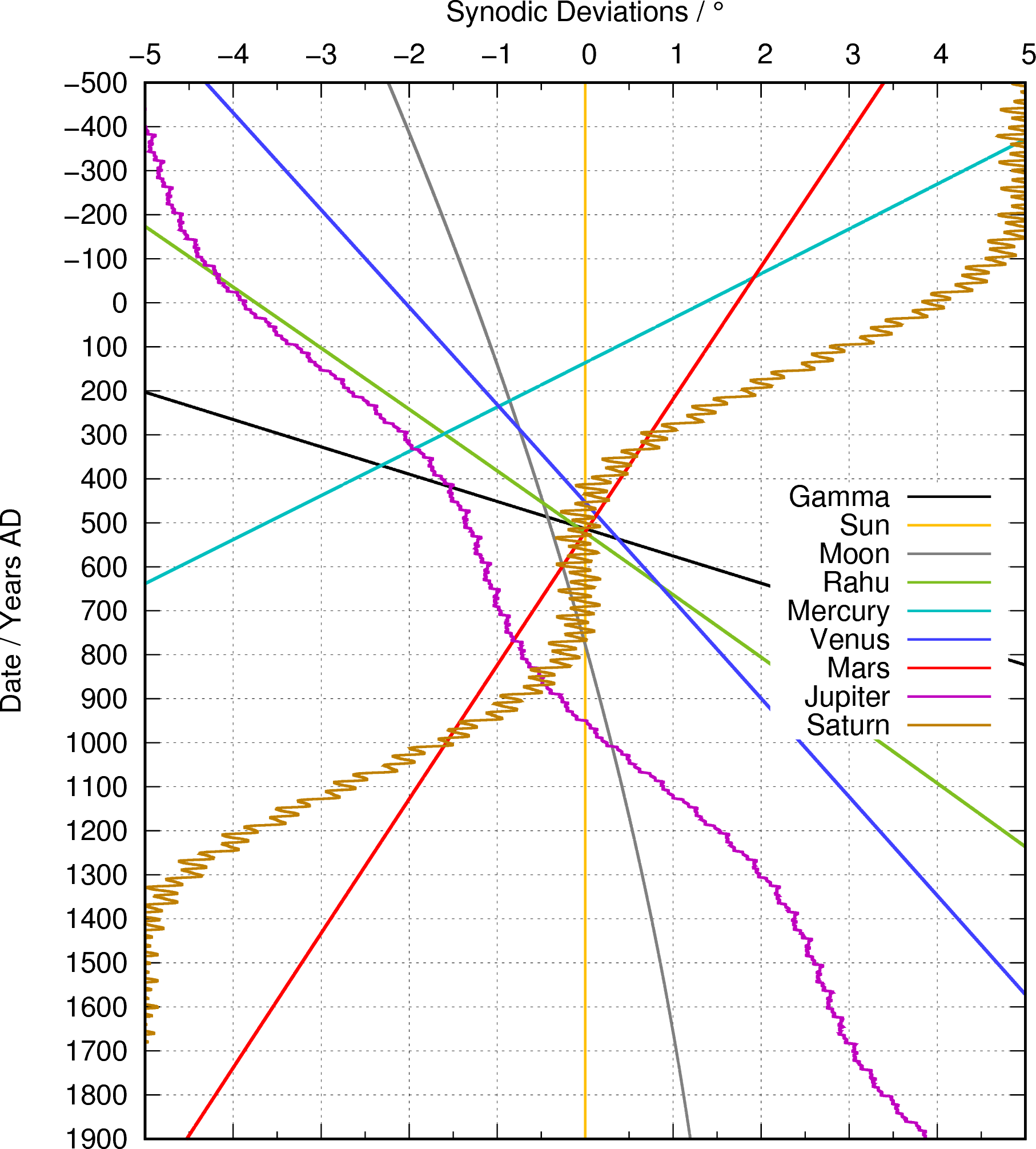}
\caption{Mean longitude deviations (left) and synodic deviations (right) for all luminaries from 500 BC to 1900 AD.\label{fig:meanlo_synodic}}
%\end{figure}

\vspace{6mm}

%\begin{figure}
\includegraphics[width=6.2cm]{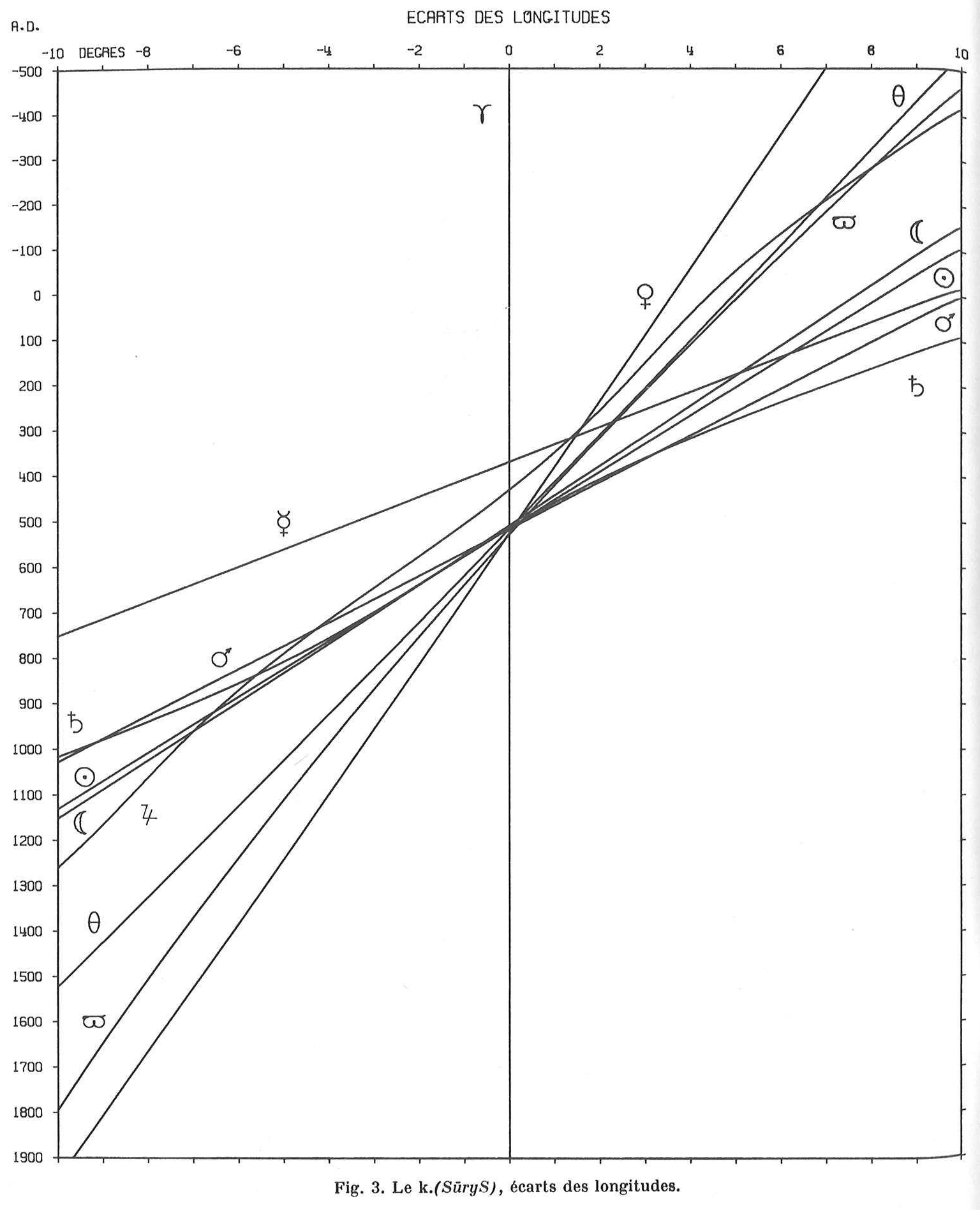} \hfill \includegraphics[width=6.2cm]{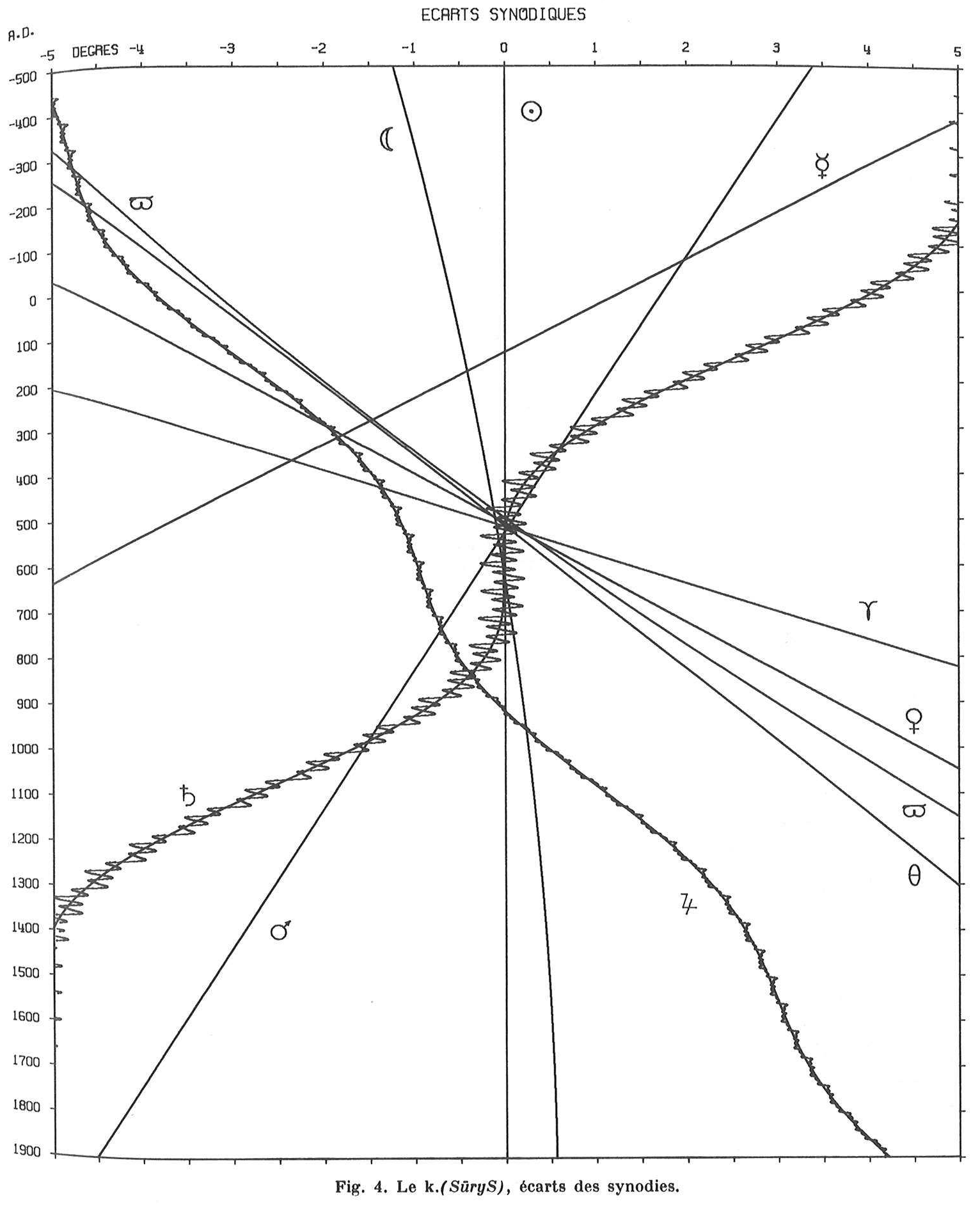}
\caption{Mean longitude deviations (left) and synodic deviations (right) plotted by Billard (see \cite{billard1971}, Fig. 3 and 4. pp. 184--185).\label{fig:fig3_4_billard}}
\end{figure}

In order to have a first overview of this problem, we have plotted the graphs of the mean longitude deviations and of the synodic deviations (see Fig. \ref{fig:meanlo_synodic}). The tightening of the deviations in the vicinity of 500 AD is evident. For the mean longitude deviations, this effect is mainly due to the precession of the equinoxes. On the other hand, for the synodic deviations, it comes from a lack of adjustment of the planetary motion constants $\left\{\alpha_i\right\}$. In any case, this tightening points out the time frame during which %**
 observations were carried out to determine these constants $\left\{\alpha_i\right\}$ and $\left\{\beta_i\right\}$. Nevertheless, we can observe that the behavior of Mercury and Jupiter diverges from the other luminaries.

The second point of interest of these graphs concerns their extreme resemblance with those plotted by Billard (see Fig. \ref{fig:fig3_4_billard}) in the analysis of the ephemerides described in (the older version of) the \textit{S\={u}rya Siddh\={a}nta}, around 500 AD (see \cite{billard1971} pp. 73--83).
The similarity of these graphs provides further evidence for an Indian origin of the methods of computing ephemerides. However, we notice a slight discrepancy between the trajectory of the moon in the synodic deviation graphs of Fig.~\ref{fig:meanlo_synodic} and \ref{fig:fig3_4_billard}. The moon being the fastest luminary, its position is most sensitive to a change in the terrestrial longitude of the observation site. Thus, this difference may reflect an adaptation of the Indian ephemerides to another geographical location.

% Compare mean longitude deviations + synodic deviations with Fig. 3 and 4 of Billard

	\subsection{Direct method on the mean longitude deviations}
We have applied the direct method on the mean longitude deviations described in \S \ref{sec:direct_method} to different sets of luminaries. 
 In this case, the mean longitude deviation of the Vernal point cannot be used since it is identically null. 
 Recall that the reference terrestrial longitude of observation was set here to 90$\deg$ East, i.e. -90$\deg$, so that, for example, $\Delta\phi_0=-14\deg$ corresponds to 104$\deg$ East (of Greenwich). %**
 The luminary positions were computed from 500 BC to 2000 AD. However, since this method relies on the linearity of the luminary trajectories around their passage at the origin, we have also used this method by limiting the period within 200 years around 500 AD.

		\subsubsection{Results of the computation over the period [500 BC -- 2000 AD]\label{sec:500_2000}}
\begin{tabular}{lll}
(01101\,11111) & $t_0 = $ 470 AD $\pm$ 80 y & $\Delta \phi_0=$ -14 $\pm$ 100$\deg$ \\
(01101\,01101) & $t_0 = $ 524 AD $\pm$ 19 y & $\Delta \phi_0=$ -19 $\pm$ 20$\deg$ \\
(01101\,01100) & $t_0 = $ 516 AD $\pm$ 22 y & $\Delta \phi_0=$ -15 $\pm$ 19$\deg$ \\
(01101\,01001) & $t_0 = $ 528 AD $\pm$ 27 y & $\Delta \phi_0=$ -20 $\pm$ 25$\deg$ \\
(01101\,00101) & $t_0 = $ 522 AD $\pm$ 24 y & $\Delta \phi_0=$ -19 $\pm$ 24$\deg$ \\
(01100\,01101) & $t_0 = $ 526 AD $\pm$ 22 y & $\Delta \phi_0=$ -20 $\pm$ 23$\deg$ \\
(01101\,00000) & $t_0 = $ 509 AD $\pm$ 100 y & $\Delta \phi_0=$ -13 $\pm$ 64$\deg$
\end{tabular}

\noindent
The confidence intervals are given at 95 \%.

		\subsubsection{Results of the computation over the period [300 AD -- 700 AD]\label{sec:300_700}}
\begin{tabular}{lll}
(01101\,01101) & $t_0 = $ 518 AD $\pm$ 13 y & $\Delta \phi_0=$ -12 $\pm$ 13$\deg$ \\
(01101\,01100) & $t_0 = $ 518 AD $\pm$ 20 y & $\Delta \phi_0=$ -12 $\pm$ 18$\deg$ \\
(01101\,01001) & $t_0 = $ 520 AD $\pm$ 20 y & $\Delta \phi_0=$ -12 $\pm$ 17$\deg$ \\
(01101\,00101) & $t_0 = $ 516 AD $\pm$ 6 y & $\Delta \phi_0=$ -12 $\pm$ 6$\deg$ \\
(01100\,01101) & $t_0 = $ 519 AD $\pm$ 16 y & $\Delta \phi_0=$ -12 $\pm$ 16$\deg$ \\
(01101\,00000) & $t_0 = $ 513 AD $\pm$ 52 y & $\Delta \phi_0=$ -10 $\pm$ 34$\deg$
\end{tabular}

\noindent
The confidence intervals are given at 95 \%.

		\subsubsection{Comments on the results of the direct method}
The first observation concerns the compatibility of all confidence intervals, both for $t_0$ and for $\Delta \phi_0$. We see that $t_0$ seems to be between 500 and 530. Only one estimate is quite different, 470, but with a huge uncertainty domain: 80 years!

On the other hand, the accuracy of the results are much more accurate by limiting the period to [300 AD -- 700 AD]. This is particularly clear for $\Delta \phi_0$ whose confidence intervals are significantly narrower.

Theoretically, the more luminaries we use, the better the accuracy. However, if some luminaries have erroneous constants $\alpha_i$ and/or $\beta_i$, they will degrade the estimation of $(t_0,\Delta \phi_0)$. This is what happened in the first line of \S \ref{sec:500_2000} where all luminaries are used. The use of Mercury and Jupiter increases drastically both uncertainties on $t_0$ as well as on $\Delta \phi_0$.

In \S \ref{sec:300_700}, Mercury and Jupiter have been omitted. However, the best estimation is still not the one using the more luminaries, (01101\,01101), but the 4$^\textrm{\footnotesize th}$: (01101\,00101). It seems that avoiding Venus improves drastically the estimates of $t_0$ and $\Delta \phi_0$. For this method, we will retain these confidence intervals:
$$
t_0=516\pm 6 \quad \textrm{and} \quad \Delta \phi_0=-12\pm 6\deg \quad \textrm{@ 95 \% confidence.}
$$

	\subsection{Variance method on the synodic deviations}
We have applied the variance method on the synodic deviations described in \S \ref{sec:variance_method} to different set of luminaries. In this case, the Sun cannot be used since its synodic deviation is identically null. However, the synodic deviation of the Vernal point can be used thanks to equation (\ref{eq:gamma}).

		\subsubsection{Results of the variance method}
The following results have been obtained by varying the date from 300 to 700 BC and the terrestrial longitude from -45$\deg$ to +45$\deg$ around the meridian 90$\deg$ East.
\begin{tabular}{lll}
(10101\,11111) & $t_0 = $ 415 AD $\pm$ 160 y & $\Delta \phi_0=$ 0 $\pm$ 100$\deg$ \\
(10101\,01101) & $t_0 = $ 520 AD $\pm$ 17 y & $\Delta \phi_0=$ -7 $\pm$ 8$\deg$ \\
(10101\,01100) & $t_0 = $ 518 AD $\pm$ 63 y & $\Delta \phi_0=$ -7 $\pm$ 35$\deg$ \\
(10101\,01001) & $t_0 = $ 525 AD $\pm$ 57 y & $\Delta \phi_0=$ -8 $\pm$ 21$\deg$ \\
(10101\,00101) & $t_0 = $ 520 AD $\pm$ 64 y & $\Delta \phi_0=$ -6 $\pm$ 31$\deg$ \\
(10100\,01101) & $t_0 = $ 518 AD $\pm$ 31 y & $\Delta \phi_0=$ -8 $\pm$ 13$\deg$ \\
(10001\,01101) & $t_0 = $ 517 AD $\pm$ 56 y & $\Delta \phi_0=$ +59 $\pm$ 230$\deg$ \\
(00101\,01101) & $t_0 = $ 518 AD $\pm$ 48 y & $\Delta \phi_0=$ -7 $\pm$ 13$\deg$
\end{tabular}

\noindent
The confidence intervals are given at 95 \%.

		\subsubsection{Comments on the results of the variance method}
Here also, the estimates are all compatible. Similarly, the use of all available luminaries is of no help: the first line shows huge uncertainties (160 years and 100$\deg$).

It may be noticed that the estimation performed without the Moon, (10001\,01101), gives a very poor uncertainty over $\Delta \phi_0$: $230\deg$! This was predictable because the Moon is the most efficient luminary to estimate $\Delta \phi_0$ since it has the shortest period.

The best estimates are obtained with all luminaries except Mercury and Jupiter (10101\,01101) and we will retain these confidence intervals, which are very close to the one of the direct method:
$$
t_0=520\pm 17 \quad \textrm{and} \quad \Delta \phi_0=-7\pm 8\deg \quad \textrm{@ 95 \% confidence.}
$$
Let us remark that the confidence interval over $t_0$ is almost three times the one of the direct method, whereas the confidence interval over $\Delta \phi_0$ is of the same order of magnitude.

\begin{figure}
\includegraphics[width=\linewidth]{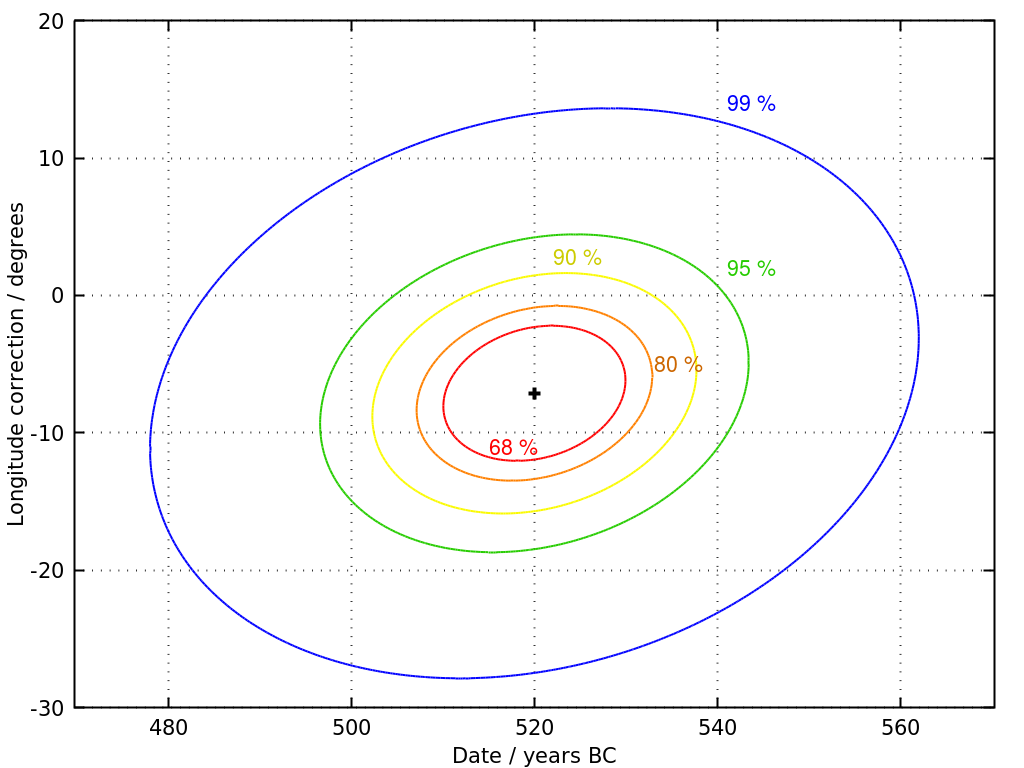}
\caption{Probability levels on $t_0$ and $\Delta \phi_0$. The black cross represents the estimates$(\hat{t}_0,\Delta \hat{\phi}_0)$ obtained by the variance method applied to (10101\,01101).\label{fig:PDF}}
\end{figure}

Let us remind the reader %**
 that the estimates of $t_0$ and $\Delta \phi_0$ are not independent but are correlated. Fig. \ref{fig:PDF} shows the probability levels of the true values $(t_0,\Delta \phi_0)$ over the estimates $(\hat{t}_0,\Delta \hat{\phi}_0)$ in the plane of $(t,\Delta \phi)$. %**
% comment on the validity of the results
% Thorough study of the case SMRVMaS 300-700 AD (figures: paraboloid map -not 3D-, t0  Dphi0 probability 2D map - 68%, 80%, 90%, 95% level curves-)

\section{Discussion\label{sec:discussion}}
	\subsection{Coincidence date}
As stated previously, the coincidence date $t_0$ points to the epoch when astronomical observations were carried out in order to determine the constants $\{\alpha_i\}$ and $\{\beta_i\}$ for each luminary. These constants are essential to compute the ephemerides for any time. Thanks to the direct method, the coincidence date $t_0$ can be set precisely at 516 AD $\pm$ 6 years. 
 It points towards the \=Aryabha\d ta epoch and demonstrates that the Khmer ephemerides described in \cite{faraut1910} are an adaptation of a version of the \textit{S\={u}rya Siddh\={a}nta}, which may have been established at the longitude of Ujjain, India (longitude 76$\deg$ East). However, this is compatible with later, complementary observations in Cambodia, performed in order to correct the \textit{S\={u}rya Siddh\={a}nta} ephemerides for the new epoch and location. %**

	\subsection{Coincidence location}
\begin{figure}
\includegraphics[width=\linewidth]{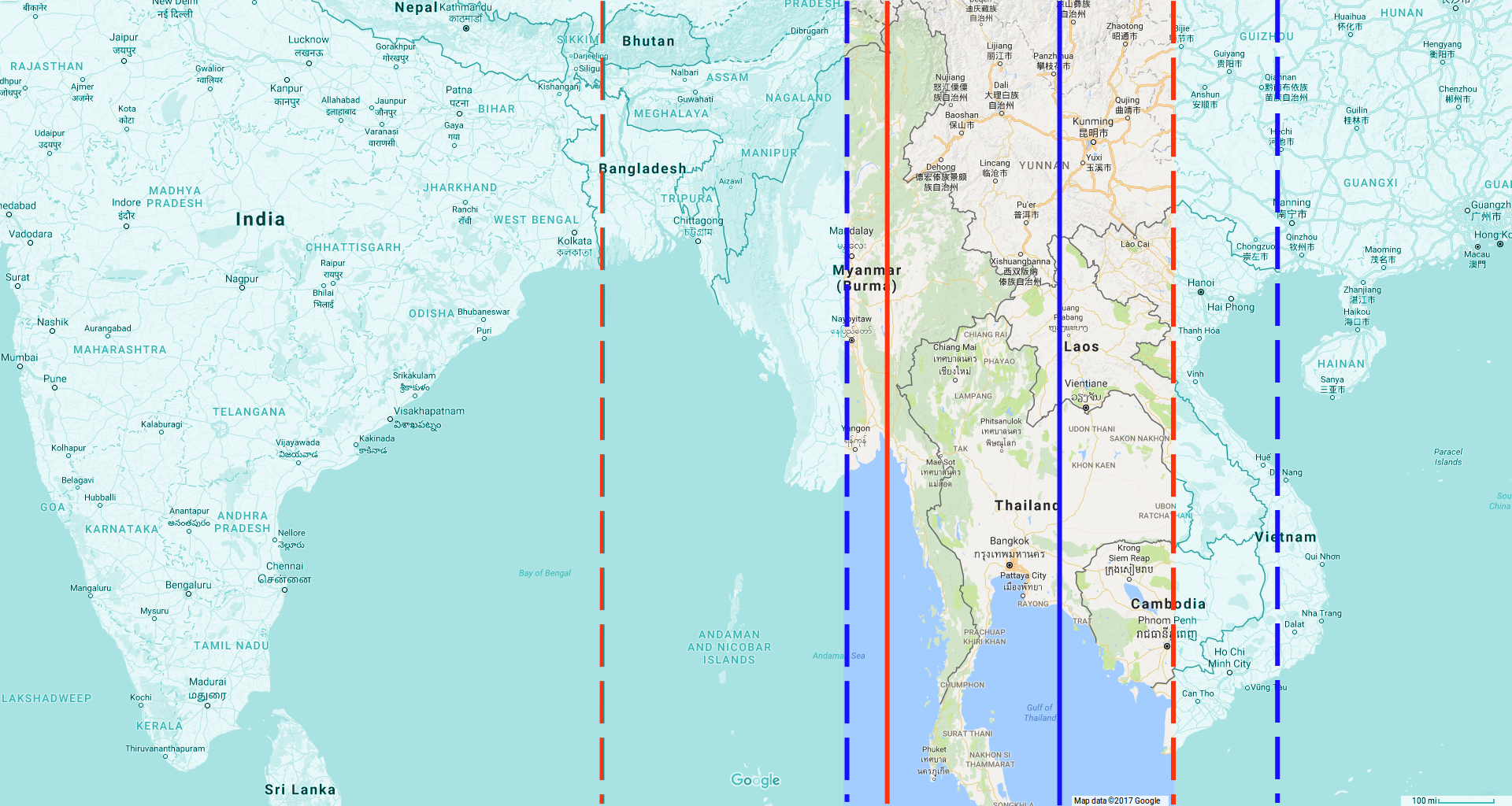}
\caption{Locations compatible with the estimate of $\Delta \phi_0$ obtained with the direct method (the 95 \% confidence interval is between the blue dashed lines) and with the variance method (the 95 \% confidence interval is between the red dashed lines). The highlighted %**
 part of the map is within both 95\% intervals.\label{fig:southeast_map}}
\end{figure}

The coincidence location longitude is given by $90\deg+\Delta \phi_0$. Thanks to our two estimates, it yields:
$$
\phi_0=-102\pm 6 \deg \quad \textrm{and} \quad \phi_0=-97\pm 8 \deg \quad \textrm{@ }95\textrm{\% confidence.}
$$
These confidence intervals point towards the vicinity of Burma and Thailand but even the least favorable is compatible with the western part of Cambodia including Angkor and Phnom Penh
 (see Fig. \ref{fig:southeast_map}).

Concerning Burma, it is interesting to report the remarks of Billard previously mentioned in footnote \ref{fn:Burma} p.~\pageref{fn:Burma}.

	\subsection{Comparison of Indian and Khmer ephemerides\label{sec:adapt_terrestrial_longitude}}
We already noticed the deep similarity between the Indian ephemerides of \textit{S\={u}rya Siddh\={a}nta} and the Khmer ephemerides in \cite{faraut1910}. According to \cite{brahmagupta628} and \cite{billard1971}, \textit{k.(SuryS)} relies on the assumption of a general conjunction of all luminaries at the origin of \textit{KYardh}, in 3102 BC. %**
 Therefore, in this canon, all the $\{\beta^0_i\}$ of the luminaries are identically null.

In order to compare these ephemerides, we first have %**
 to convert the $\{\beta^0_i=0\}$ coefficients of the luminaries with the \textit{KYardh} origin to the $\{\beta'_i\}$ coefficients with the \textit{f.638} origin. Denoting $\{\alpha'_i\}$ the mean motion of the luminary $i$ in the \textit{k.(SuryS)} ephemerides, it comes:
\begin{equation}
\alpha'_i t_{KYardh}=\alpha'_i t_{f.638} + \beta'_i.
\end{equation}
Knowing that $t_{KYardh}=t_{f.638}+1\,365\,702$ days:
\begin{equation}
\beta'_i =1\,365\,702 \times \alpha'_i.\label{eq:beta_prime}
\end{equation}

\begin{table}
\hspace{-15mm}\begin{tabular}{|l||c|r|r||c|r|}
\hline
 & $\alpha'_i$ & \multicolumn{1}{c|}{$360/\alpha'_i$} & \multicolumn{1}{c||}{Deviation} & $\beta'_i$ & \multicolumn{1}{c|}{Deviation} \\
Luminary & ratio & \multicolumn{1}{c|}{decimal} & \multicolumn{1}{c||}{from $\alpha_i$} & ratio & \multicolumn{1}{c|}{from $\beta_i$} \\
 & ($\deg/$day) & \multicolumn{1}{c|}{(days)} & \multicolumn{1}{c||}{('/millenium)} & ($\deg$) & \multicolumn{1}{c|}{(')}\\
\hline
\hline
&&&&&\\
Sun & $\displaystyle\frac{288\,000}{292\,207}$ & $365.2588$ & $0.00$ & $\displaystyle\frac{393\,322\,176\,000}{292\,207}$ & $3.00$ \\
&&&&&\\
\hline
&&&&&\\
Moon & $\displaystyle\frac{19\,251\,112}{1\,461\,035}$ & $27.32167$ & $-38.06$ & $\displaystyle\frac{26\,291\,282\,160\,624}{1\,461\,035}$ & $39.18$ \\
&&&&&\\
\hline
&&&&&\\
Lunar asc. node & $\displaystyle-\frac{232\,226}{4\,383\,105}$ & $-6\,794.751$ & $-42.58$ & $\displaystyle-\frac{1\,394\,025\,216}{1\,461\,035}$ & $-1.31$ \\
&&&&&\\
\hline
&&&&&\\
Mercury & $\displaystyle\frac{1\,195\,800}{292\,207}$ & $87.96999$ & $5.12$ & $\displaystyle\frac{1\,018\,879\,120}{292\,207}$ & $7.85$ \\
&&&&&\\
\hline
&&&&&\\
Venus & $\displaystyle\frac{2\,340\,796}{1\,461\,035}$ & $224.6982$ & $284.45$ & $\displaystyle\frac{1\,374\,110\,568}{1\,461\,035}$ & $27.41$ \\
&&&&&\\
\hline
&&&&&\\
Mars & $\displaystyle\frac{765\,608}{1\,461\,035}$ & $686.9999$ & $2.10$ & $\displaystyle\frac{1\,915\,323\,888}{1\,461\,035}$ & $0.84$ \\
&&&&&\\
\hline
&&&&&\\
Jupiter & $\displaystyle\frac{72\,844}{876\,621}$ & $4\,332.321$ & $5.36$ & $\displaystyle-\frac{1\,198\,672\,872}{292\,207}$ & $12.74$ \\
&&&&&\\
\hline
&&&&&\\
Saturn & $\displaystyle\frac{146\,564}{4\,383\,105}$ & $10\,766.07$ & $-4.54$ & $\displaystyle-\frac{1\,998\,560\,760}{1\,461\,035}$ & $1.08$ \\
&&&&&\\
\hline
\end{tabular}
\caption{Constants $\{\alpha'_i\}$ and $\{\beta'_i\}$ of the luminaries according to \textit{k.(SuryS)}. The $\{\alpha'_i\}$ constants are given in \cite{brahmagupta628} and \cite{billard1971}. The $\{\beta'_i\}$ constants are deduced from the $\{\alpha'_i\}$ thanks to (\ref{eq:beta_prime}). The deviations are calculated as $\alpha'_i-\alpha_i$ and $\beta'_i-\beta$ (i.e. Indian coefficient - Khmer coefficient).\label{tab:alpha_beta_prime}}
\end{table}

The $\{\alpha'_i\}$ and $\{\beta'_i\}$ constants obtained are very close to the $\{\alpha_i\}$ and $\{\beta_i\}$ constants of the Khmer ephemerides (see Tables \ref{tab:alpha_beta} and \ref{tab:alpha_beta_prime}).

		\subsubsection{Origin of the differences between the mean motions}
We can first %**
 notice that the mean motion of the Sun is described by exactly the same ratio in Table \ref{tab:alpha_beta} and Table \ref{tab:alpha_beta_prime}.

Except for Venus and to a lesser extent %**
 for the Moon and the lunar ascending node, % ajout de la V2
the discrepancies between the ephemerides are around 5' per millenium. For Venus however, the difference reaches almost 5$\deg$ per millenium. It seems then that the \textit{k.(SuryS)} coefficients have been kept and the only differences could be due to truncation errors between ratios. %, except for Venus,
% the Moon and the lunar ascending node. % ajout de la V2
% The period of Venus in the Khmer ephemerides is closer to the modern value of its period than the \textit{k.(SuryS)} value. It seems then that the mean motion of Venus has been improved in the Khmer ephemerides.

We can guess how the ratios of Table \ref{tab:alpha_beta} could have been deduced from those of Table \ref{tab:alpha_beta_prime}:
\begin{enumerate}
	\item compute the duration (in hours) %**)
 of the luminary period from its $\alpha_i'$ coefficient $T_i=360\times 24/\alpha_i'$
	\item round to the nearest integer hour $\hat{T}_i \in \mathbb{N}$
	\item compute the simplified mean motion $\alpha_i=360\times 24/\hat{T}_i$.
\end{enumerate}

\subsubsubsection{Example of Jupiter}
\begin{enumerate}
	\item Computation of Jupiter period $T_9$ in hours:
$$
T_9 = \frac{360\times 24}{\alpha'_9} = \frac{360\times 24\times 876\,621}{72\,844}= 103\,975.69 \textrm{ h}
$$
	\item Rounding to the hour: $103\,976$ h
	\item Computation of the approximated ratio:
$$
\alpha_9=\frac{360 \times 24}{103\,976}=\frac{8\,640}{103\,976}=\frac{1080 \times 8}{12997 \times 8}=\frac{1080}{12997} \textrm{ } \deg/\textrm{d.}
$$
\end{enumerate}
This process would have been successively applied to Mercury and Venus rounded to the 100$^\textrm{\footnotesize th}$
 of day, to R\={a}hu, Mars and Jupiter rounded to the hour and to Saturn rounded to the day. Thus, the large differences of the mean motion of Venus and R\={a}hu are only due to particularly unfavorable truncation errors.

For the Moon however, the difference comes from the very different approaches followed by %**
 these two canons: in the Khmer one, its mean motion is given relatively to the Sun and not to the Vernal point. This point is important in view of the care taken in the determination of eclipses in Cambodia. Therefore, the $\alpha_3$ and $\beta_3$ constants of the Moon seem to have a different origin than \textit{k.(SuryS)} or at least to have been corrected.

		\subsubsection{Origin of the difference between the $\{\beta_i\}$ and the $\{\beta'_i\}$ constants\label{sec:dif_beta}}
Concerning, the $\{\beta'_i\}$ compared to the $\{\beta_i\}$, it appears that the main differences for the Sun and the Moon are the corrective ratios in Table \ref{tab:alpha_beta}: $-3/60\deg$ for the Sun and $-40/60\deg$ for the Moon. These corrections appears to be corrections of terrestrial longitude. For the Sun:
\begin{equation}
-\frac{3}{60}=\alpha_2 \frac{\Delta\phi}{360} \quad \Rightarrow \quad \Delta\phi=-18\deg\label{eq:correc_sun}
\end{equation}
where $\alpha_2$ is the mean motion of the Sun. Similarly, for the Moon:
\begin{equation}
-\frac{40}{60}=\alpha_3 \frac{\Delta\phi}{360} \quad \Rightarrow \quad \Delta\phi=-18\deg\label{eq:correc_moon}
\end{equation}
where $\alpha_3$ is the mean motion of the Moon. It is interesting to remark that Cassini comes to the same conclusion in \cite{la_loubere1691} (see Tome second, Réflexions sur les règles Indiennes, \S I, pp.~191--194).

Therefore, these corrections are intended to use the ephemerides $18\deg$ East of Ujjain, i.e. $94\deg$ East. This correction could be the one mentioned by Billard (``\textit{ce kara\d na est entré en usage, avec une excellente correction de méridien, en basse Birmanie}'', see footnote \ref{fn:Burma} p. \pageref{fn:Burma}) and is fully compatible with our confidence intervals over $\phi_0$. However, this correction is not appropriate for any location in Cambodia.%**

The deviations $\beta'_i-\beta_i$ of the other luminaries range from 1 to 25' and do not seem to be related either to the corresponding $\alpha_i$ or to the corrective terms in Table \ref{tab:alpha_beta}, unlike those shown in (\ref{eq:correc_sun}) and (\ref{eq:correc_moon}). %**
 It may be noticed that they all are positive and then all induce a $\Delta \phi$ to the East (except for the lunar ascending node but since its coefficient $\alpha_5$ is negative, it points also to the East). %**).

\section{Conclusion\label{sec:conclusion}}
The Khmer ephemerides reported in \cite{faraut1910} are undoubtedly closely related to \textit{k.(SuryS)}, the Indian ephemerides described in the \textit{S\={u}rya Siddh\={a}nta}. 
According to Billard, the latter could represent lost work by \=Aryabha\d ta, and were elaborated around 510 AD at the longitude of Ujjain, 76$\deg$ East. In his introduction to \cite{burgess1935} (pp. xxxv--xliii), Sengupta points out that the modern version of the \textit{S\={u}rya Siddh\={a}nta} that we have nowadays may be much more recent (between 400 and 1100 AD), and contains decisive input from other astronomers such as Brahmagupta (b.~598). %**
 Nevertheless, Billard's work as well as Mercier's work confirm this version of the canon was undoubtedly elaborated around 510 AD at the Ujjain longitude.

According to our study, the Khmer ephemerides were most likely based on observations performed around 510 AD: they present almost exactly the same deviations from modern ephemerides. It is then clear that the Khmer ephemerides are an adaptation of the \textit{k.(SuryS)} for another location.%***
 The only marks of longitude correction clearly presented by the Khmer ephemerides are the subtraction of 3' for the Sun and 40' for the Moon. The corrective terms of the other luminaries are much more difficult to interpret but could come from an attempt at %***
 observational rectification of the canon. Moreover, such corrections are suitable for 94$\deg$ East.
Except for the Moon, it seems probable that the ephemerides described by Faraut are nothing but the adaptation of \textit{k.(SuryS)} for Burma, probably in the seventh century, confirming Billard \cite[p.~74]{billard1971}. %***
 However, they do not seem to be identical with the Parahita system described by Haridatta \cite{parahita}, 
 because the mean parameters in the Khmer canon are much closer to those of \textit{k.(SuryS)} that to Haridatta's.

While this model was imported into Cambodia at a much later date, this nevertheless confirms that elaborate astronomical knowledge was nurtured outside India at that time. The famous first attestation of zero as a digit\footnote{It is not the place to discuss the evidence for the existence of this system earlier in India. We merely recall that Brahmagupta explained in 628 AD how to extend the rules of integer arithmetic to negative numbers and zero, as well as fractions and quadratic surds. In his case, zero is not just a digit, it is actually a number.} (\ in inscription K. 127 in seventh century Cambodia\cite{coedes}, together with the numbering of months from 0 to 11 rather than 1 to 12 \cite[p.~16]{faraut1910} suggest that Indian learning, already well-known to have influenced Cambodia in many ways, may also have been nurtured in its scientific dimensions, possibly along original lines. Since we know that there was a partial break in the continuity of Indian Mathematics, between 628 AD and the ninth century \cite{sk-2010,sk-2012}, our results open the tantalizing possibility that key information on the evolution of mathematical ideas in India, unavailable from those Indian sources that are still extant, might be gathered from the study of mathematics and astronomy in Cambodia.

We hope that this work will encourage an examination of earlier ephemerides in Cambodia and more generally, of the interplay between mathematical concepts and other aspects of the sciences and of culture in this country.

\section*{Acknowledgements}

We are grateful to Professor Raymond Mercier for his valuable help.

\bibliography{Khmer_ephemerides_v6.bib}

\section*{Appendix: Use of true longitudes}
\renewcommand{\thesubsection}{A.\arabic{subsection}}
	\subsection{Computation of the true longitudes}
\begin{figure}
\hspace{-6.5mm} \includegraphics[height=7.6cm]{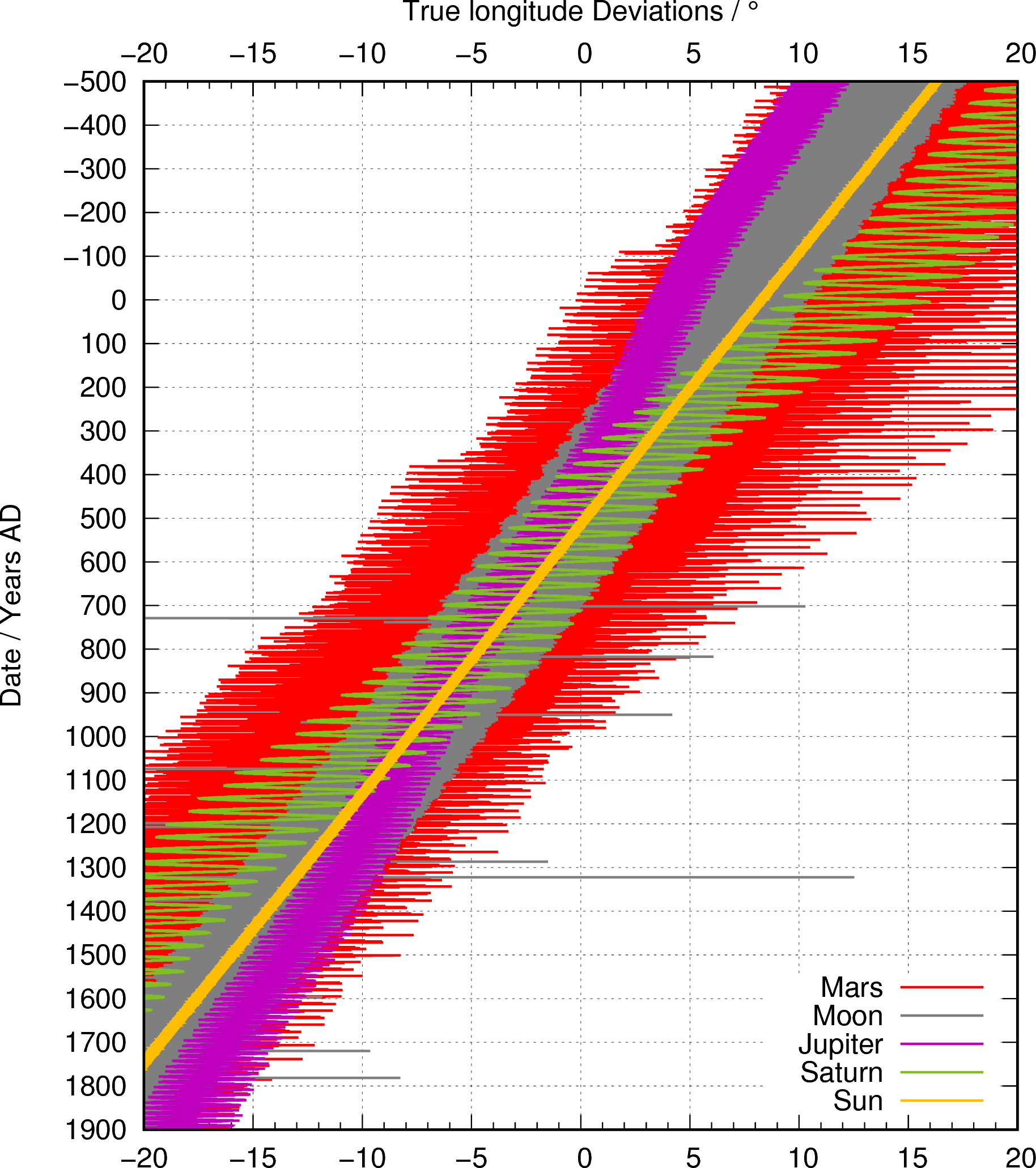} \includegraphics[height=7.6cm]{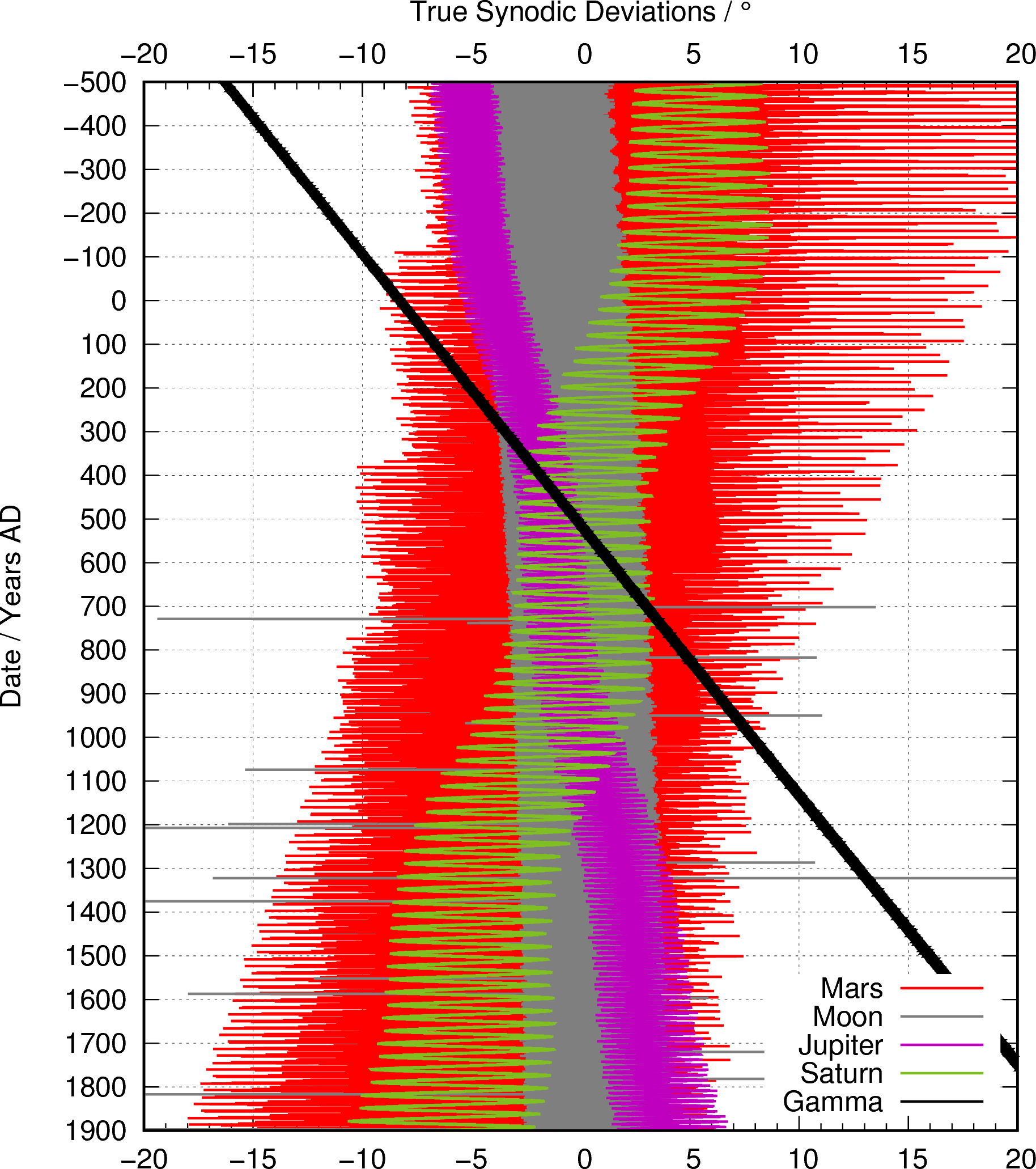}
\caption{True longitude deviations (left) and synodic deviations (right) for the Vernal Point, the Sun, the Moon, Mars, Jupiter and Saturn from 500 BC to 1900 AD.\label{fig:truelo_synodic}}
\end{figure}

As stated above, we began this study by using the true longitudes instead of the mean longitudes. At the Khmer true longitudes computed according to Faraut \cite{faraut1910}, we subtract the  VSOP87D ephemerides \cite{bretagnon_francou1988}. Figure \ref{fig:truelo_synodic} presents the true longitude deviations of several luminaries for the period 500 BC to 1900 AD.

The first thing that strikes is the enormous dispersion of the trajectory of each luminary except for the Sun. It appears that true longitudes are affected by periodic errors at relatively short periods (a few days). Mercury and Venus were not represented because their dispersion was such that they made these graphs illegible. These errors may come from transcription errors in \cite{faraut1910} (several were detected), from bugs in our codes as well as from internal errors of the canon itself. The magnitude of this dispersion is such that we can only put limited confidence in the following results.

We used the direct method described in \S \ref{sec:direct_method}. The terrestrial longitude of observation was set to 90$\deg$ East and the luminary position were computed from 500 BC to 2000 AD and over the limited period within 200 years around 500 AD.

	\subsection{Results of the computation over the period [500 BC -- 2000 AD]\label{sec:A500_2000}}
\begin{tabular}{lll}
(01100\,11111) & $t_0 = $ 510 AD $\pm$ 46 y & $\Delta \phi_0=$ -12 $\pm$ 56$\deg$ \\
(01100\,01101) & $t_0 = $ 523 AD $\pm$ 20 y & $\Delta \phi_0=$ -18 $\pm$ 21$\deg$ \\
(01100\,01100) & $t_0 = $ 515 AD $\pm$ 3 y & $\Delta \phi_0=$ -14 $\pm$ 3$\deg$ \\
(01100\,01001) & $t_0 = $ 526 AD $\pm$ 35 y & $\Delta \phi_0=$ -20 $\pm$ 33$\deg$ \\
(01100\,00101) & $t_0 = $ 525 AD $\pm$ 33 y & $\Delta \phi_0=$ -19 $\pm$ 33$\deg$ \\
\end{tabular}

\noindent
The confidence intervals are given at 95 \%.

	\subsection{Results of the computation over the period [300 AD -- 700 AD]\label{sec:A300_700}}
\begin{tabular}{lll}
(01100\,01101) & $t_0 = $ 517 AD $\pm$ 1.8 y & $\Delta \phi_0=$ -11 $\pm$ 1.8$\deg$ \\
(01100\,01100) & $t_0 = $ 516 AD $\pm$ 2 y & $\Delta \phi_0=$ -11 $\pm$ 2$\deg$ \\
(01100\,01001) & $t_0 = $ 517 AD $\pm$ 4 y & $\Delta \phi_0=$ -11 $\pm$ 3$\deg$ \\
(01100\,00101) & $t_0 = $ 517 AD $\pm$ 3 y & $\Delta \phi_0=$ -11 $\pm$ 3$\deg$ \\
\end{tabular}

\noindent
The confidence intervals are given at 95 \%.

	\subsection{Comments}
These results are fully consistent with the ones obtained with the mean longitudes. The confidence interval are even significantly narrower for the date as well as for the terrestrial longitude. However, because of the errors mentioned above, we will not retain these results.

\end{document}